\documentclass[fleqn,usenatbib]{mnras}

\usepackage{newtxtext,newtxmath}

\usepackage[T1]{fontenc}

\DeclareRobustCommand{\VAN}[3]{#2}
\let\VANthebibliography\thebibliography
\def\thebibliography{\DeclareRobustCommand{\VAN}[3]{##3}\VANthebibliography}

\usepackage{graphicx}	
\usepackage{amsmath}	

\title[A 3D map of local \ion{H}{I} emission]{Milky Way Atlas: A radial-velocity-resolved, three-dimensional map of \ion{H}{I} within 1.25~kpc}

\author[Lewis McCallum et al.]{
Lewis McCallum,$^{1}$\thanks{E-mail: mccallum@physik.rwth-aachen.de}
Laurin Söding,$^{2}$
Mareike Berkner,$^{1}$
Philipp Frank,$^{3}$
Matteo Guardiani,$^{2}$
Jakob Roth$^{4}$,
\newauthor
Juan D. Soler,$^{5}$
Robert Benjamin,$^{6}$
Torsten En\ss lin,$^{2,7,8,9}$
Philipp Mertsch,$^{1}$
Konstantinos Tassis,$^{10,11}$
\newauthor
Vasiliki Pavlidou$^{10,11}$
\\
$^{1}$Institute for Theoretical Particle Physics and Cosmology, RWTH Aachen University, Sommerfeldstraße 16, 52074 Aachen, Germany\\
$^{2}$Max Planck Institute for Astrophysics, Karl-Schwarzschild-Str. 1, 85748 Garching, Germany\\
$^{3}$Kavli Institute for Particle Astrophysics \& Cosmology, P.O. Box 2450, Stanford University, Stanford, CA 94305, USA\\
$^{4}$Max Planck Computing and Data Facility, Gie\ss enbachstraße 2, 85748 Garching, Germany\\
$^{5}$University of Vienna, Department of Astrophysics, Türkenschanzstra\ss e 17, 1180 Wien, Austria\\
$^{6}$University of Wisconsin, Whitewater, Whitewater, WI 53190, USA\\
$^{7}$Deutsches Zentrum für Astrophysik, Postplatz 1, 02826 Görlitz, Germany\\
$^{8}$Ludwig-Maximilians-Universit\"at M\"unchen, Geschwister-Scholl-Platz 1, 80539 Munich, Germany\\
$^{9}$Excellence Cluster ORIGINS, Boltzmannstr. 2, 85748 Garching, Germany\\
$^{10}$University of Crete, Department of Physics,  Institute of Theoretical \& Computational Physics, \& Institute of Astrobiology, 70013 Herakleio, Greece\\
$^{11}$Institute of Astrophysics, Foundation for Research and Technology-Hellas, 71110 Heraklion, Crete, Greece
}

\date{Accepted XXX. Received YYY; in original form ZZZ}

\pubyear{\the\year{}}

\begin{document}
\label{firstpage}
\pagerange{\pageref{firstpage}--\pageref{lastpage}}
\maketitle

\begin{abstract}
We present a velocity-resolved three-dimensional map of local atomic hydrogen (\ion{H}{I}) within 1.25~kpc of the Sun, tackling the challenge of converting emission from position--position--velocity space into true 3D structure. Our method combines the HI4PI full-sky survey with the Edenhofer et al. (2024) 3D dust map in the framework of Information Field Theory, enabling a joint reconstruction of the local \ion{H}{I} density, radial velocity field, and effective line width while also separating emission arising inside the mapped local volume from more distant Galactic \ion{H}{I}. The inference is driven by morphological matching between dust and \ion{H}{I} structures together with kinematic coherence in 3D space. Synthetic data tests show that the method recovers the local density and velocity structure, even in the presence of substantial contamination from distant emission. The resulting map reveals a smoother, more diffuse local \ion{H}{I} distribution than the dust, a declining \ion{H}{I}-to-dust ratio toward high dust column densities consistent with the atomic-to-molecular transition, and a velocity field that captures both large-scale Galactic rotation and local non-circular velocities. Independent comparisons with maser and young stellar cluster velocities agree with the recovered kinematics. This \ion{H}{I} map provides a new three-dimensional, kinematically resolved view of the nearby atomic interstellar medium and a foundation for localising other velocity-resolved Galactic emission in physical space.
\end{abstract}

\begin{keywords}
ISM: structure -- ISM: kinematics and dynamics -- ISM: clouds -- ISM: dust, extinction -- radio lines: ISM -- Galaxy: kinematics and dynamics
\end{keywords}



\section{Introduction}

Atomic hydrogen (\ion{H}{I}) is one of the fundamental constituents of the interstellar medium (ISM), tracing the diffuse gas from which molecular clouds form and through which stellar feedback propagates \citep{mccluregriffiths23}. Its 21~cm hyperfine transition provides an all-sky view of Galactic structure, with modern surveys such as HI4PI delivering high angular and spectral resolution observations across the full sky \citep{hi4pi}. However, \ion{H}{I} observations measure the ISM in position--position--velocity (PPV) space rather than true three-dimensional position space, and converting line emission into physical distance remains a longstanding challenge in studies of Galactic \ion{H}{I}, recognised already in pioneering work by \citet{westerhout57} and \citet{oort58}. In the Milky Way, multiple structures frequently overlap along the same line of sight, while kinematic distances derived from Galactic rotation are subject to degeneracies and departures from the assumed circular motion \citep{kalberla09}. As a result, even in the nearby Galaxy, the physical placement of \ion{H}{I} structures remains uncertain.

Recent advances in 3D dust mapping have provided additional information. Using stellar photometry and parallaxes from \emph{Gaia}, a series of dust reconstructions have mapped the nearby ISM in 3D with ever improving spatial resolution and accuracy \citep{leike20,lallement22}. Most recently, \citet{edenhofer23} produced a parsec-scale map of differential dust extinction out to 1.25~kpc from the Sun, based on extinctions derived by \citet{zhang23}. Dust extinction measurements carry direct distance information through the underlying stellar parallaxes, making such maps a powerful anchor for locating interstellar structures in physical space. While dust and atomic gas do not trace identical phases of the ISM, their morphology is closely related, and correlations between dust and \ion{H}{I} have been observed both in emission and extinction \citep{planck11,planck14}. This makes dust a natural choice for localising \ion{H}{I} emission in 3D.

A number of studies have begun to bridge the gap between spectroscopic gas surveys and distance-resolved ISM structure. A commonly used benchmark for the vertical distribution of Galactic \ion{H}{I} is the profile compiled by \citet{dickey90} from earlier 21-cm studies of the inner Milky Way, particularly \citet{lockman84}. In that work, \ion{H}{I} emission at terminal velocities in the inner Galaxy was used to isolate gas near the tangent points, where the geometry of the rotating disc provides a well constrained location for the emitting material. This made it possible to infer how the mean \ion{H}{I} density varies with height above and below the Galactic plane, and \citet{dickey90} summarised the result in a simple analytic form that remains widely used as a reference for the Galactic \ion{H}{I} layer.

More recently, kinetic tomography approaches have combined distance-resolved dust information with spectroscopic gas tracers to infer ISM velocity structure, beginning with the work of \citet{kt1,kt2}. Several other applications have focused on specific regions of the sky, including nearby clouds and selected local-ISM fields \citep{ivanova21,duchene23,soler23}. In particular, \citet{ivanova21} and \citet{duchene23} use absorption features in stellar spectra, such as \ion{K}{I}, to associate velocities with 3D dust structures. These absorption-line tracers provide complementary information to 21-cm \ion{H}{I} emission: they give velocity information tied to stars with distance estimates, but they probe particular species and physical conditions rather than the full atomic hydrogen distribution. As a result, structures seen in \ion{K}{I} absorption need not correspond directly to the same velocity components or column-density features seen in \ion{H}{I} \citep{nguyen25}. More recently, \citet{soler25} applied the histogram of oriented gradients (HOG) method \citep{soler19} to morphologically associate PPV \ion{H}{I} and CO emission with distance-resolved dust structures in the Galactic plane, thereby inferring the local non-circular gas velocity field.

In parallel, Bayesian field-inference methods have begun to produce 3D reconstructions of gas emission on larger scales \citep{mertsch21,mertsch23,soding25}. Together, these developments demonstrate both the constraining power of dust tomography for localising ISM structures in distance, and the ability of Information Field Theory (IFT; \citet{ensslin09}) based reconstructions to incorporate 3D spatial correlations as a prior. However, an IFT-based, velocity-resolved \ion{H}{I} reconstruction directly tied to the newest generation of high-resolution dust maps remains absent. A fundamental difficulty is that much of the observed \ion{H}{I} emission along any line of sight originates outside the volume for which precise 3D dust information is available, introducing a degeneracy between local gas and more distant emission.

In this work we combine the HI4PI full-sky 21~cm survey with the 3D dust map of \citet{edenhofer23} within the framework IFT \citep{ensslin09}. We treat the unknown \ion{H}{I}-to-dust ratio, radial velocity and line-width as smooth 3D fields, use a forward model to predict the HI4PI data they would produce, and then infer which field configurations best match the observations while remaining spatially correlated and physically plausible according to our set of priors. Distance information enters through the dust map, which constrains dense regions in 3D space via the 54 million \emph{Gaia} parallaxes used in its construction. We infer the local \ion{H}{I}-to-dust ratio, radial velocity field, and effective line width, while introducing a remainder component that absorbs emission not originating within the local 1.25~kpc volume. The high dimensionality of this problem is handled using Metric Gaussian Variational Inference (MGVI) \citep{knolmuller19}, implemented in the \texttt{NIFTy} framework \citep{selig13,steininger19,edenhofer24_nifty}.

At an intuitive level, this method works due to a combination of morphological matching between dust and gas structures, and 3D velocity coherence which also matches the data. By morphological matching, it is meant that when a structure exists both in the sky projection of the dust map, as well as the gas data, the optimiser is able to adjust the 3D \ion{H}{I}-to-dust ratio in that 3D volume to match the intensity of gas emission on the sky, as well as localising it in velocity space (as is informed by the data). The velocity coherence helps in regions where morphological matching is less constraining. This is that gas which is less well constrained by morphological matching can still be well constrained by being kinematically coherent with other, better constrained gas emission. Structures in the gas emission data which cannot be explained under either morphological matching or velocity coherence without moving too far from the prior, are explained at lower cost in the remainder. This is how we are able to break the local/distant degeneracy.

The result is a velocity-resolved three-dimensional map of local atomic hydrogen within 1.25kpc of the Sun. We validate the method through synthetic data tests that include realistic distant contamination, and compare the recovered velocity structure to masers and young stellar clusters with measured line-of-sight (LOS) velocities.

\section{Data and Methods}

\subsection{Data}

Our methodology relies heavily on both the \ion{H}{I} emission data from \citet{hi4pi}, and the 3D dust map of \citet{edenhofer23}.

\subsubsection{HI4PI}

The HI4PI survey is the highest spatially and spectrally resolved full-sky dataset probing the neutral atomic hydrogen in the Milky Way. The dataset is an assimilation of observations from the Effelsberg–Bonn \ion{H}{I} Survey \citep{kerp11,winkel16} and the Galactic All-Sky Survey (GASS) \citep{mccluregriffiths09,kelberla10,kalberla15}, and gives the brightness temperature of the 21cm \ion{H}{I} emission line as a function of radial velocity for the full sky, sampled on an $N_{\rm side} = 1024$ HEALPix grid \citep{gorski05}. We use an $N_{\rm side}=64$ version of this dataset. The degradation from $N_{\rm side} = 1024$ to $N_{\rm side} = 64$ is done by averaging the values in all child pixels independently for each velocity channel. We also seek to probe only the local ISM ($d<1.25~\rm kpc$), and thus truncate the spectral axis of this dataset to $|v_{\rm LSR}| < 75~ \rm km ~s^{-1}$. Within a distance of $1.25~\rm kpc$, the highest expected velocities due to Galactic rotation are around $20~\rm km~s^{-1}$ \citep{reid19}. This allows for non-circular velocities of up to $55~\rm km~s^{-1}$ throughout the volume. We find through testing that our results do not change by including data from larger velocity ranges. The full spectral resolution of the dataset is used, with velocity resolution of $\Delta V_{\rm LSR} \approx 1.288~\rm km~s^{-1}$.

\subsubsection{Edenhofer Dust Map}

The dust map of \citet{edenhofer23} is a 3D map of differential extinction due to dust out to a distance of $1.25~\rm kpc$ from the Sun. This map was constructed using 54 million stars with \emph{Gaia} parallaxes and extinction values determined by \citet{zhang23} using \emph{Gaia} BP/RP spectra. \citet{edenhofer23} modelled the distribution of dust extinction as a Log-Normal Gaussian process, resulting in a map which probes dust structures in 3D at parsec-scale resolution. The \citet{edenhofer23} map was retrieved on the relevant grid using the \texttt{dustmaps} python package \citep{dustmaps}.

The \citet{edenhofer23} dust map is provided in units of $E~\rm pc^{-1}$, where $E$ is the extinction unit introduced by \citet{zhang23}. To convert this quantity to a dust mass density, we proceed in a sequence of steps using the extinction curve of \citet{zhang23} and the astrodust$+$PAH Milky Way $R_V=3.1$ model of \citet{hensley23}. Although extinction is produced by dust, converting extinction to dust mass requires an assumed dust opacity. Because the \citet{hensley23} opacity is tabulated per hydrogen nucleus, $\tau_\lambda/N_{\rm H}$, we use hydrogen column density as an intermediate quantity and then apply the model gas-to-dust ratio to obtain a dust mass.

First, \citet{zhang23} provide the conversion from $E$ to monochromatic extinction, $A_\lambda/E$. For a given wavelength $\lambda$, this gives
\begin{equation}
\frac{dA_\lambda}{ds}
=
\left(\frac{A_\lambda}{E}\right)\frac{dE}{ds}.
\end{equation}

Second, we convert extinction in magnitudes to optical depth using the standard relation
\begin{equation}
A_\lambda = 1.086\,\tau_\lambda,
\end{equation}
so that
\begin{equation}
\frac{d\tau_\lambda}{ds}
=
\frac{1}{1.086}\frac{dA_\lambda}{ds}.
\end{equation}

Third, \citet{hensley23} tabulate the extinction cross section per hydrogen nucleus, $\tau_\lambda/N_{\mathrm H}$, for the adopted dust model. Dividing the optical-depth density by this quantity gives the hydrogen number density,
\begin{equation}
n_{\mathrm H}
=
\frac{d\tau_\lambda/ds}{\tau_\lambda/N_{\mathrm H}}.
\end{equation}

Finally, we convert this to a dust mass density. Using the total gas-to-dust mass ratio of 140 (including helium) from \citet{hensley23}, the corresponding gas mass density is $1.4\,m_{\mathrm H}n_{\mathrm H}$, and hence
\begin{equation}
\rho_{\mathrm{dust}}
=
\frac{1.4\,m_{\mathrm H}}{140}\,n_{\mathrm H}.
\end{equation}

Combining these steps gives
\begin{equation}
\rho_{\mathrm{dust}}
=
\frac{1.4\,m_{\mathrm H}}{140}
\frac{1}{1.086}
\left(\frac{A_\lambda}{E}\right)
\left(\frac{dE}{ds}\right)
\left(\frac{\tau_\lambda}{N_{\mathrm H}}\right)^{-1}.
\end{equation}

We find that the conversion factor derived is nearly constant with respect to wavelength chosen between $400$--$2200\,\mathrm{nm}$. We therefore adopt
\begin{equation}
\rho_{\mathrm{dust}} \simeq 4.5 \times 10^{-23}
\left(\frac{dE/ds}{E ~pc^{-1}}\right)\,
\mathrm{g\,cm^{-3}}.
\end{equation}
Here, $dE/ds$ is the differential extinction value from \citet{edenhofer23}.

We note that our derived conversion  of $4.5\times10^{-23}\rm~g~cm^{-3}$ per dust map unit is somewhat higher than in previous works which convert the \citet{edenhofer23} dust map to 3D hydrogen densities (such as \citet{zucker21,oneill24,mccallum25}). In these works, total hydrogen number density is obtained with a conversion factor of $1652 \rm ~cm^{-3}/E~pc^{-1}$. The equivalent conversion using our updated method can be found by multiplying our $4.5\times10^{-23}\rm~g~cm^{-3}$ by the \citet{hensley23} hydrogen-to-dust ratio of 100, and converting mass density to hydrogen number density. The result is $2700 \rm ~cm^{-3}/E~pc^{-1}$.

This difference is due to updates in the dust model between \citet{draine09} and \citet{hensley23}. \citet{zucker21} uses figure~1 of \citet{draine09} to obtain a conversion from extinction to hydrogen column density at $\lambda=673~\rm nm$, and finds a value of $A_{673}/N_{H} = 4.0~\rm mag~cm^{-2}$. The models of \citet{hensley23} however give $A_{673}/N_{H} = 2.43~\rm mag~cm^{-2}$, accounting for this $\approx1.6$ factor increase in dust and hydrogen density.

\subsubsection{Masers and Young Stellar Clusters}
\label{vel_tracers}

In order to gauge the success of our gas velocity reconstruction, we also include a comparison of our final map to two datasets. We first use the positions and radial velocities of local masers from \citet{reid19}. We expect the maser velocities to trace the velocity of the gas very closely, as the observations detect masing molecules which are actually part of the gas. While these objects represent the cleanest data to compare our reconstructed velocities to, they come with relatively large parallax errors as well as large velocity measurement errors. There are also only 11 of these sources within our volume of interest.

We also include the dataset of young stellar clusters from \citet{hunt23}. These objects are expected to trace the gas less directly than the masers, since young clusters can drift from their birth clouds and may already have partially decoupled from the surrounding ISM. The observable used in our comparison is the catalogue mean stellar radial velocity of each cluster, derived from Gaia DR3 radial velocities of cluster members. We use only clusters with ages $<10~\mathrm{Myr}$ according to \citet{hunt23}, of which 207 lie within 1.25~kpc in the parent sample. We then apply the quality cuts recommended by \citet{hunt23}; 123 clusters pass these additional cuts within 1.25~kpc.

For both the masers and young clusters, we evaluate the reconstructed gas LOS velocity at the object's sky position and distance, sampling over the quoted parallax uncertainty, and compare this to the observed LOS velocity. For the clusters, this observed velocity is the mean stellar radial velocity of the member stars, while for the masers it is the measured maser radial velocity.

\subsection{Methods}

This work follows the broad philosophy of kinetic tomography initiated by \citet{kt1}, who pioneered the combination of 3D dust/reddening information with \ion{H}{I} and CO velocity data to infer a four-dimensional ISM distribution in longitude, latitude, distance, and radial velocity. Our reconstruction inherits this idea, but improves upon it in several respects. Most notably, we benefit from modern high-fidelity 3D dust reconstructions, using the parsec-scale \citet{edenhofer23} map, and from the full-sky HI4PI \ion{H}{I} survey. In this sense, the present work is a next-generation realisation of kinetic tomography in the local ISM, with substantially enhanced angular and distance resolution. Methodologically, we formulate the reconstruction in the framework of Information Field Theory (IFT) \citep{ensslin09,ensslin19}, using correlated Gaussian-process priors, the ICR representation of fields on HEALPix grids \citep{edenhofer22}, and Metric Gaussian Variational Inference \citep{knolmuller19}, implemented in \texttt{NIFTy} \citep{selig13,steininger19,edenhofer24_nifty}. This lets us infer continuous 3D fields for the local \ion{H}{I} density, LOS velocity, and line width, while simultaneously fitting a flexible remainder component for emission outside the reconstructed volume.

\subsubsection{Grid Properties}

We seek to reconstruct the \ion{H}{I} emission within the same volume as is covered by the \citet{edenhofer23} dust map. This is a spherical volume which reaches out to 1.25~kpc in each direction from the Sun. The dust map contains no 3D information on the volume within a radius of 69~pc from the Sun, and we similarly will not reconstruct this volume. The grid structure will be similar to that used by \citet{edenhofer23} and \citet{soding25}. We use a HEALPix discretisation on the sky and a set of radial bins. These bins define spherical shells between 69 and 300 pc that are linearly spaced in radius; beyond 300 pc, the radial spacing becomes logarithmic. We use an $N_{\rm side}=64$ HEALPix structure for the sky axis, and 196 radial bins. Each HEALPix sky contains $12\times64^{2} = 49,152$ pixels, giving us a total of $9,633,792$ 3D voxels. In this grid setup, our 3D resolution changes as a function of radius, and is different in the plane of sky versus radial axis. To best illustrate the resolution of our grid, we include table~\ref{resotable}, which shows effective resolutions for various radius ranges.

\begin{table}
\centering
\caption{Effective spatial resolution of our 3D grid. Here $d_\perp$ is the transverse (sky-plane) resolution, taken as the HEALPix pixel chord length at the centre of the radial bin, and the aspect ratio is $dr/d_\perp$. Values reported in this table are averages through each given radius range.}
\begin{tabular}{lccc}
\hline\hline
Radius [kpc] & $dr$ [pc] & $d_\perp$ [pc] & Aspect ratio \\
\hline
$0.07$--$0.10$ & 3.36 & 1.54 & 2.20 \\
$0.10$--$0.30$ & 3.36 & 3.60 & 1.03 \\
$0.30$--$0.50$ & 4.39 & 7.08 & 0.62 \\
$0.50$--$0.80$ & 7.17 & 11.56 & 0.62 \\
$0.80$--$1.25$ & 11.27 & 18.17 & 0.62 \\
\hline
\end{tabular}
\label{resotable}
\end{table}

\subsubsection{Variational Inference}

Our goal is to find possible states of the ISM which are consistent with the observed data \citep{hi4pi} and assumed priors. To probe the very high-dimensional probability distribution of possible consistent states, we rely on Variational Inference (VI). Specifically, we use the VI method of Metric Gaussian Variational Inference (MGVI) \citep{knolmuller19}  which approximates the posterior with a high-dimensional Gaussian. MGVI finds the best possible approximation by minimizing the Kullback-Leibler (KL) divergence between the true and the approximate posterior. Furthermore, MGVI draws samples from the approximating Gaussian to probe the variance, representing the uncertainty of the posterior distribution. These variational methods are implemented in the \texttt{NIFTy} Python package \citep{selig13,steininger19,edenhofer24_nifty}.

Further work has been done to improve the sampling of such high-dimensional posteriors, such as the development of geometric variational inference (geoVI) \citep{frank21}. While MGVI approximates the posterior as a high-dimensional Gaussian, geoVI applies a non-linear transform (informed by the posterior landscape geometry as probed by the Fisher information metric) which makes the posterior much closer to a Gaussian in the transformed parameter space. While for many large scale reconstructions this has proved to be a crucial tool, we found through testing (with both synthetic and real data) that this extra step had little impact on the final results of our field inference. Because of this, we avoid the extra computational expense of geoVI, and use MGVI as our variational inference method.

Throughout this paper we will refer to behaviour of the VI optimiser in a number of ways, one of which is to describe a move in parameter space as being `low-cost' for the optimiser. This means that it is able to make that move in parameter space while staying comfortably within regions of large prior probability (priors described in section~\ref{priorsection}). We will also often refer to parameters as having been `learned' by the optimiser, by which we mean the MGVI optimiser is moving this parameter (by moving the mean of the Gaussian distribution) in a certain direction in order to explain the data under the given priors.

\subsubsection{The Forward Model}

In order to evaluate the likelihood of any one ISM state given the observed HI4PI dataset, we must construct a forward model which converts a 3D ISM state into a synthetic dataset. For the VI framework to sample the posterior, this model must contain as much of the relevant physics as possible, while also running fast enough to be computationally feasible with MGVI (typically on the order of seconds or less).

To create a synthetic HI4PI dataset, we require three 3D fields: the \ion{H}{I} emitting gas density, the gas radial velocity, and the effective line width. Using these three values on a 3D grid, we can model each voxel as producing a Gaussian emission line which is itself attenuated by the intervening gas between that voxel and the observer (at the Sun). These three grids of values together define the effective ISM state that we seek to infer from the HI4PI data. The radial velocity and effective line-width grids are each modelled as a 3D correlated field in the form of a Gaussian process. The \ion{H}{I} emitting gas densities are derived from both the \citet{edenhofer23} dust map, and a third correlated field. This third field can be considered the 3D structure of the \ion{H}{I}-to-dust ratio. This construction allows the dust map to provide the baseline spatial distribution of material, while the inferred ratio field captures local departures from a fixed \ion{H}{I}-to-dust conversion.

Gas at different distances along the same sightline can have different velocities in our model. However, each individual voxel still has only one Gaussian line profile, with one central velocity and one line width. The reconstruction therefore cannot split two velocity components that occupy the same resolved 3D volume element; it represents them through a single velocity and line width. This is a remaining limitation shared with earlier kinetic-tomography approaches, but is addressed somewhat by the much higher resolution of our reconstruction than previous work such as \citet{kt1}.

For a given sky pixel \(p\) and velocity channel \(v\), the forward model predicts
\begin{equation}
T_{\rm model}(p,v)
=
\sum_i
T_{\rm spin}\,
\bigl(1-e^{-\tau_i^{\rm cell}(p,v)}\bigr)\,
e^{-\tau_{<i}(p,v)}
+
T_{\rm rem}(p,v),
\label{eq:measurement}
\end{equation}
where the sum runs over radial voxels along the line of sight. Here $\tau_{i}^{\rm cell}$ is the optical depth of voxel $i$, $\tau_{<i}$ is the cumulative optical depth of the foreground material between the observer and voxel $i$, and $T_{\rm rem}$ is the remainder-sky component. We model the voxel optical depth as
\begin{equation}
\tau_i^{\rm cell}(p,v)
=
C_{21}\,
\frac{n_i(p)\,\Delta r_i}{T_{\rm spin}\,\sigma_i(p)}
\exp\!\left[
-\frac{(v-v_i(p))^2}{2\sigma_i^2(p)}
\right],
\label{eq:taucell}
\end{equation}
with $C_{21} = 679.168$. Folded into this constant is the conversion from number density to brightness temperature, as well as a scaling between the CGS density units and kpc unit of $\Delta r_{i}$, the physical voxel depth. Here $n_{i}$, $v_{i}$, and $\sigma_{i}$ are respectively the local \ion{H}{I} number density, LOS velocity, and effective line width in voxel $i$.

In this form, the density field sets the optical-depth amplitude, the velocity field sets the line centroid, and the line-width field sets the Gaussian width, while emission from each voxel is attenuated by the cumulative foreground optical depth.

We generate our Gaussian random fields using the method of Iterative Charted Refinement (ICR) \citep{edenhofer22}, which is a computationally efficient method of generating such fields on HEALPix grids. Each field is assigned a Matérn covariance structure \citep{matern60}, which controls its typical fluctuation amplitude, correlation length, and degree of smoothness. In practice, this specifies the power spectrum of the Gaussian random field through a small set of hyperparameters, that is an overall variance, a maximum correlation length, and an effective spectral slope. These hyperparameters are themselves inferred by the VI optimiser.

This covariance structure provides the spatial regularisation of the reconstruction. In \citet{kt1}, this was done through explicit smoothing terms that penalised large velocity differences between neighbouring voxels. Here the same role is played by the correlated-field prior: the reconstruction can retain small-scale structure where supported by the data, but still favours smooth solutions in poorly constrained regions.

The opacity of the gas to the 21~cm line is taken into account using the same method as \citet{soding25}. For each sky pixel, the intensity is evaluated along the full line of sight, with emission from more distant voxels attenuated by the optical depth of the foreground gas. As in \citet{soding25}, we assumed a fixed value of $T_{\rm spin} = 200~ \rm K$ for the 21~cm transition throughout the reconstruction. 

By modelling these quantities as independent Gaussian processes, we give the model the flexibility to create realistic correlated astrophysical environments. In practice, each field is generated from a set of standardised latent variables drawn from independent standard normal distributions (“white-noise” latent parameters), which are then transformed into spatially correlated fields. The VI framework adjusts these latent parameters to find field realisations that best explain the HI4PI data under the assumed priors.

\subsubsection{The Remainder Sky}
\label{remaindersection}

One fundamental difficulty in this method is the dominance in the HI4PI dataset of structures which originate from outside the $1.25~\rm kpc$ radius volume in which we have a high-resolution 3D dust map. To account for this, we introduce a `remainder' field, which is generated in PPV space, and represents the emission which does not originate from within the $1.25~\rm kpc$ volume. This does introduce a possible degeneracy that emission can be either explained via elevated gas-to-dust ratios within the volume, or structure in the remainder sky, however the coherent structures in the velocity and line width maps contain extra information which helps to constrain local gas. Additionally, we use a physically motivated prior for the mean value of the \ion{H}{I}-to-dust ratios, based on the canonical gas-to-dust ratio used in \citet{hensley23}.

The remainder sky is generated as one independent 2D Gaussian process for each data velocity channel. These fields are then smoothed along the velocity axis with a single Gaussian kernel. The width of this Gaussian kernel is also learned by the optimiser. This setup gives us sky-like correlations in the spatial axis, and smoother spectral correlations along the velocity axis.

To give the remainder sky object the flexibility to recover the expected large-scale latitude structure of non-local emission, we also allow it to learn a disc-like morphology in Galactic latitude. This ensures the remainder object can easily absorb emission which is not coherent with the local structure. It is crucial that the remainder object has enough flexibility, without which it becomes low-cost for the optimiser to explain data by erroneously painting in high density gas within the local volume. 

The disc-like structure applied to the remainder object takes the form of 2 exponential discs, one thin and one thick. We control the structure of the discs with a series of periodic functions of longitude, each function modelled as 1D Gaussian-smoothed field. These functions are also learned by the optimiser. 

Here the terms `thin' and `thick' disc refer only to angular latitude components in the PPV remainder model, not to a physical three-dimensional decomposition of the Galactic \ion{H}{I} disc. The remainder is defined entirely in data space, as a function of sky position and velocity, and is not assigned distances. The scale heights below are therefore angular scale heights in Galactic latitude, measured in degrees, rather than physical vertical scale heights in pc or kpc.

A reasonable set of priors for these angular scale heights was chosen by fitting the same model to the distant-sky contribution predicted by the full-Galaxy gas reconstruction of \citet{soding25}, after removing the emission within the local 1.25~kpc volume.

The unstructured correlated field in data space is multiplied by the following 2-disc structure:

\begin{equation}
a(b,\ell)=\!e^{-\frac{|b-B_0(\ell)|}{H_{\rm thick}(\ell)}}+w_{\rm thin}\,e^{S_{\rm thin}(\ell)}e^{-\frac{|b-B_0(\ell)|}{H_{\rm thin}(\ell)}}
\end{equation}

where $B_0(\ell)$, $H_{\rm thick}(\ell)$, $H_{\rm thin}(\ell)$, and $S_{\rm thin}(\ell)$ define the smooth longitude-dependent fields. They represent respectively the small latitude shift away from 0 of the discs, the scale height of the thick disc, the scale height of the thin disc and the intensity of the thin disc relative to the thick disc. $w_{\rm thin}$ is a learned scalar representing the \emph{total} intensity of the thin disc relative to the thick disc, regardless of longitude.

\subsubsection{Priors}
\label{priorsection}

Our methods relies on the generation of three 3D correlated Gaussian random fields (GRFs): the logarithm of the \ion{H}{I}-to-dust density ratio field; the LOS velocity field; and the logarithm of the line-width field. The \ion{H}{I} density field is derived by multiplying the resulting \ion{H}{I}-to-dust density field by the \citet{edenhofer23} dust map. The three GRFs share the same grid structure, but can have different correlation structures through independent learnable Matérn correlation kernels. The Matérn kernel is characterised in our implementation by the following two parameters: the average log-log slope of the power spectrum, and the turnover length-scale above which correlations are suppressed. We allow the optimiser to learn these values for each of our three 3D fields.

These correlated field priors play a role similar to the spatial smoothing terms used by \citet{kt1}, which penalises velocity differences between neighbouring voxels. In our case the regularisation is probabilistic rather than imposed as a smoothing penalty. This allows small-scale structure where supported by the data, but the reconstructed fields should still be interpreted as smooth fields: in weakly constrained regions, some of the apparent velocity smoothness may be inherited from the prior rather than uniquely required by the \ion{H}{I} data.

The \ion{H}{I}-to-dust density and line-width fields have positivity enforced by modelling them as log-normal fields; that is, they are calculated as $e^{\rm GRF}$, where GRF is the signed Gaussian random field. The velocity field is allowed to be positive or negative, and is generated as its own GRF, to which an approximate Galactic rotation curve is added. We use a flat rotation curve assuming a Solar orbital radius of 8.2~kpc, and a rotation speed of 230 $\rm km ~s^{-1}$. This only needs to be an approximation of the real rotation curve, as any departures from this assumption can be learned by the additive velocity GRF. The mean values and variances of these three fields are also learned by the VI optimiser.

The remainder sky is also constructed using the Matérn kernel, with the same correlation structure parameters of average slope and length-scale. These values are also learned, but are shared through all velocity channels. We also learn the width of Gaussian smoothing in the velocity axis as a single scalar. Also learned are the series of periodic longitude functions (as described in section~\ref{remaindersection}), each of which carries a base value, an amplitude, and a longitude smoothing width.

A summary of our used priors is shown in table~\ref{priorstable}.

\subsubsection{Synthetic Data Test}
\label{mocktestdescribe}

We carry out an end-to-end synthetic data test of our method of reconstructing the gas velocities, densities and line widths. This test must also include large contributions to the data from regions outwith the reconstructed volume. To do so, we generate Gaussian random fields as draws from our prior to represent a synthetic \ion{H}{I}-to-dust ratio, LOS velocity and line-width structure. As in the real data reconstruction, the \citet{edenhofer23} dust map is taken as a ground truth in the synthetic data test. To include the distant sky contributions, we use the full galaxy 3D gas map from \citet{soding25}. This full galaxy map without the gas at $d <\rm  1.25~kpc$ is passed through our forward model machinery in order to create an approximation of contributions to our data from distant sources. This is then added to the synthetic data generated from the three GRFs, and this creates our synthetic data product.

The synthetic data product is given to the same code as carries out our full reconstruction, and we test for success in decomposing the local and distant skies, and also for the valid reconstruction of the full 3D gas structure, gas velocity and line widths.

\subsubsection{Initial Velocity Seeding}
\label{seedsection}

\begin{figure}
    \centering
    \includegraphics[width=1.0\columnwidth]{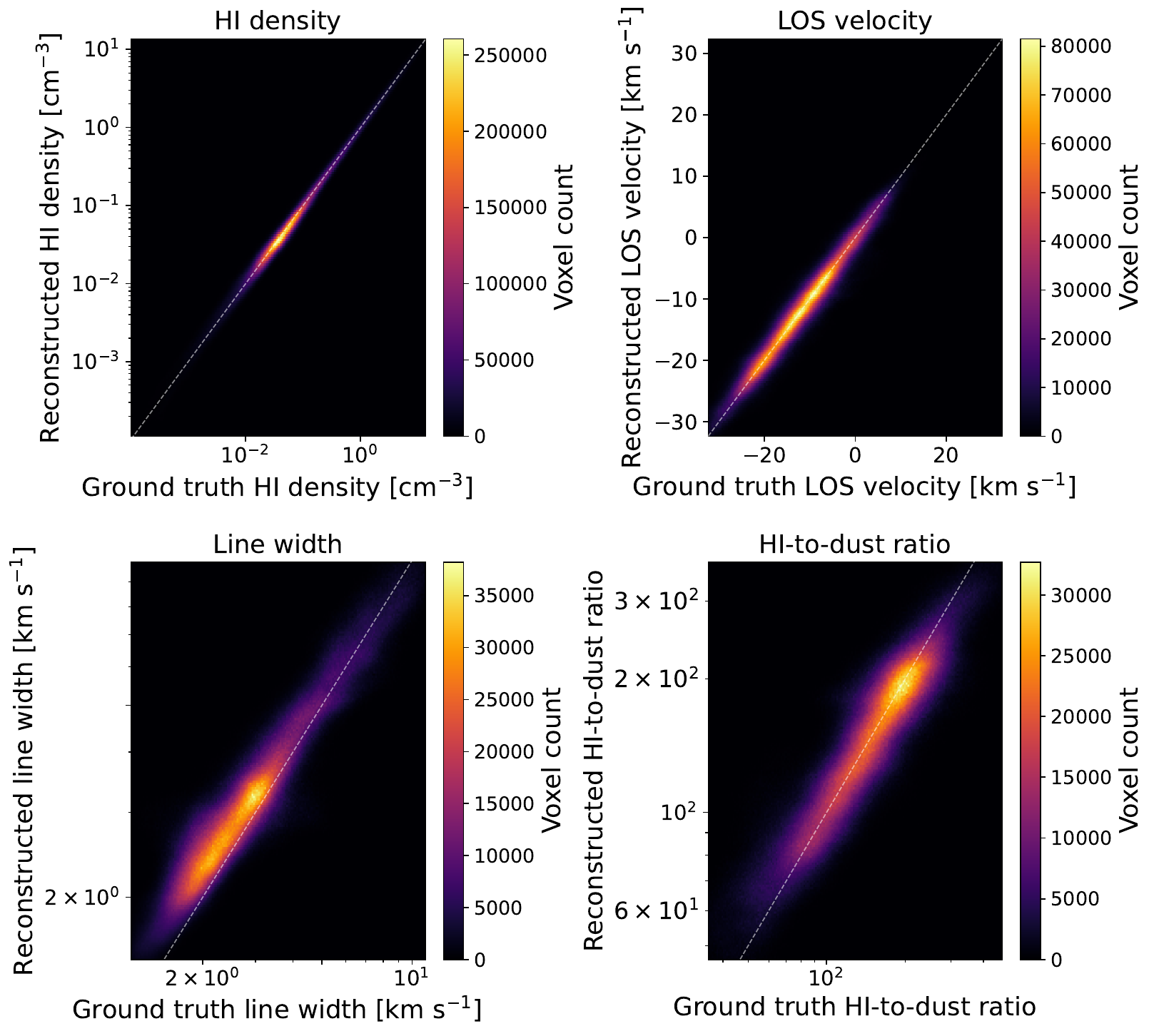}
    \caption{Comparison between our synthetic data test ground truth and reconstructed values. These 2D histograms count Cartesian voxels and show the 1:1 perfect match line. The top left panel shows the \ion{H}{I} density, the top right panel shows the radial velocity, the bottom left panel shows the line width, and the bottom right panel shows the \ion{H}{I}-to-dust mass ratio.}
    \label{fig:mockhistos}
\end{figure}

\begin{figure*}
    \centering
    \includegraphics[width=1.0\textwidth]{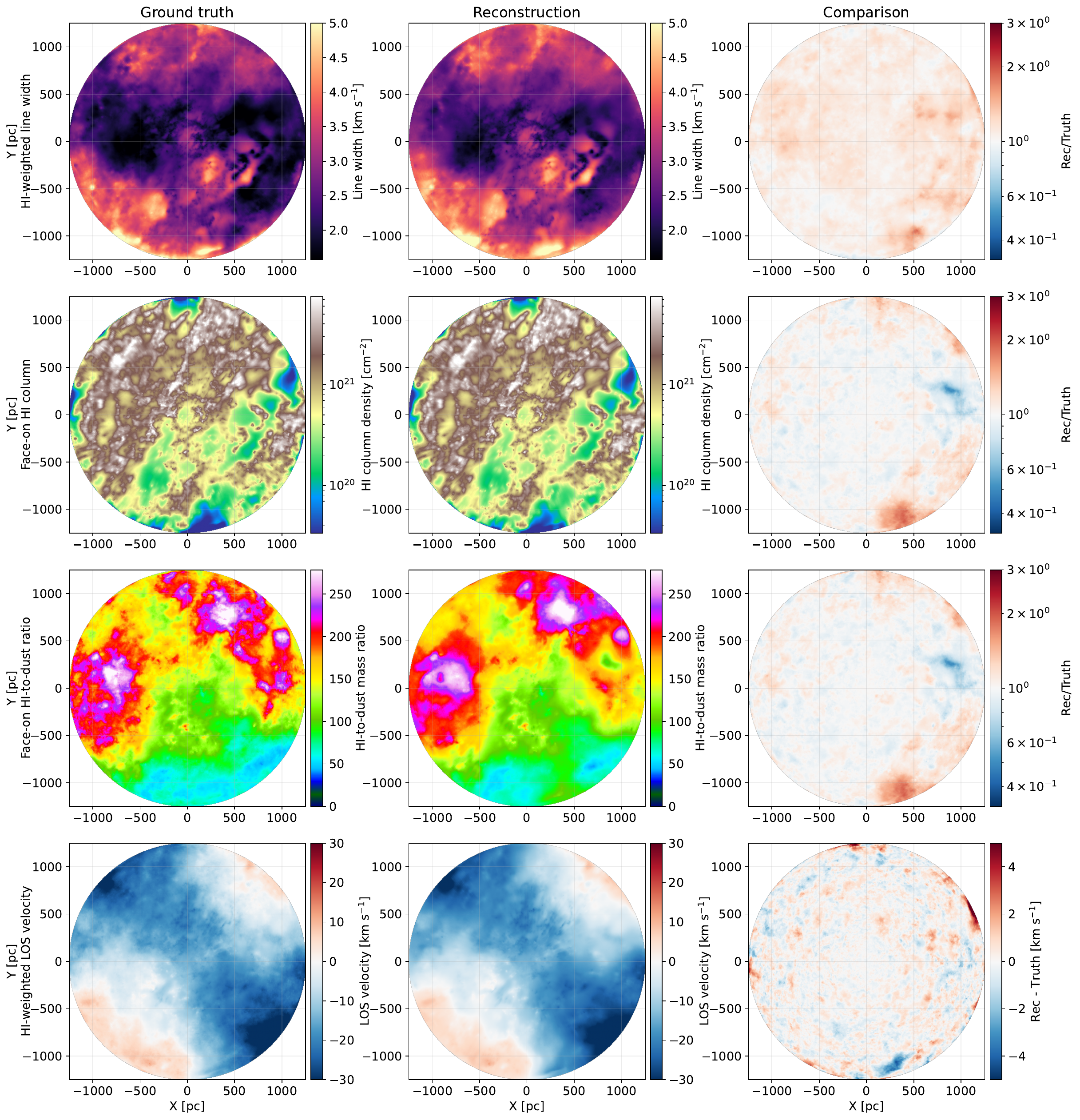}
    \caption{Top-down maps comparing our ground truth and posterior mean reconstructed maps from synthetic data. In all maps the Galactic centre is to the right ($+x$), and the direction of Galactic rotation is up ($+y$). Left columns are the truth maps, middle columns are the reconstructed maps and the right column is the difference. For all but the velocity, the difference map is a ratio between reconstructed and truth. For the velocity map, it is the difference between reconstructed and truth. The top row is the map of \ion{H}{I}-density weighted line width, the second row is the total \ion{H}{I} density, the third row is the view of \ion{H}{I}-to-dust mass ratio and the bottom row is the \ion{H}{I} density weighted LOS velocity. These maps show an excellent match but with subtle differences towards the Galactic plane, seen to the right of the top-down view of \ion{H}{I} density, \ion{H}{I}-to-dust ratio and LOS velocity.}
    \label{fig:mockmain}
\end{figure*}

\begin{figure*}
    \centering
    \includegraphics[width=1.0\textwidth]{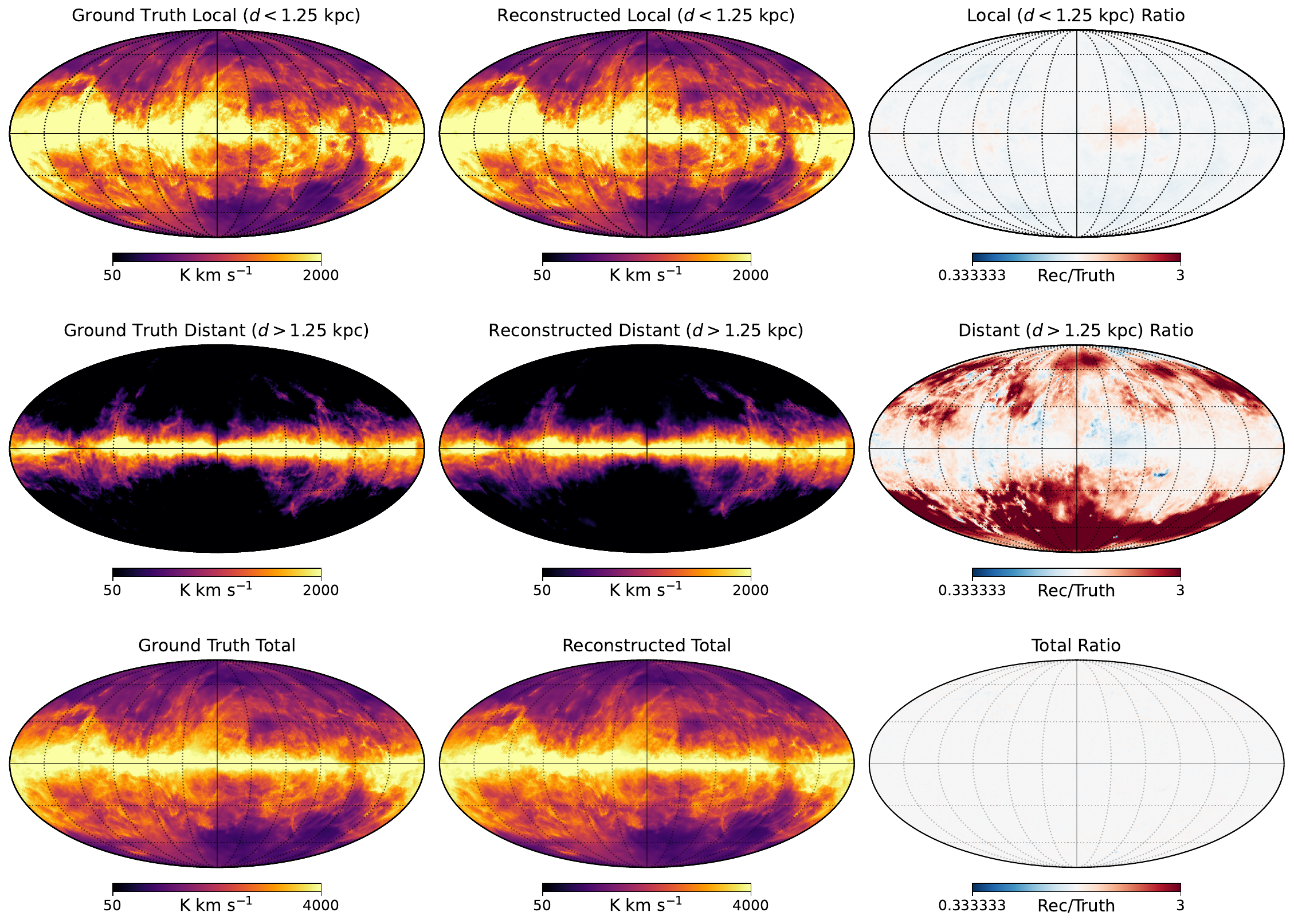}
    \caption{Comparison of ground truth to reconstruction showing total intensities on the plane of the sky, with each panel centred on the Galactic centre ($\ell=0^{\circ}$). The left column is ground truth, middle is reconstructed posterior mean, and right is the ratio between reconstructed and truth. Top row is the local only sky, middle row is the distant only sky, and the bottom row is the total reconstructed data. The definition of distant and local here is beyond and within 1.25~kpc respectively.}
    \label{fig:mockskies}
\end{figure*}

\begin{figure*}
    \centering
    \includegraphics[width=1.0\textwidth]{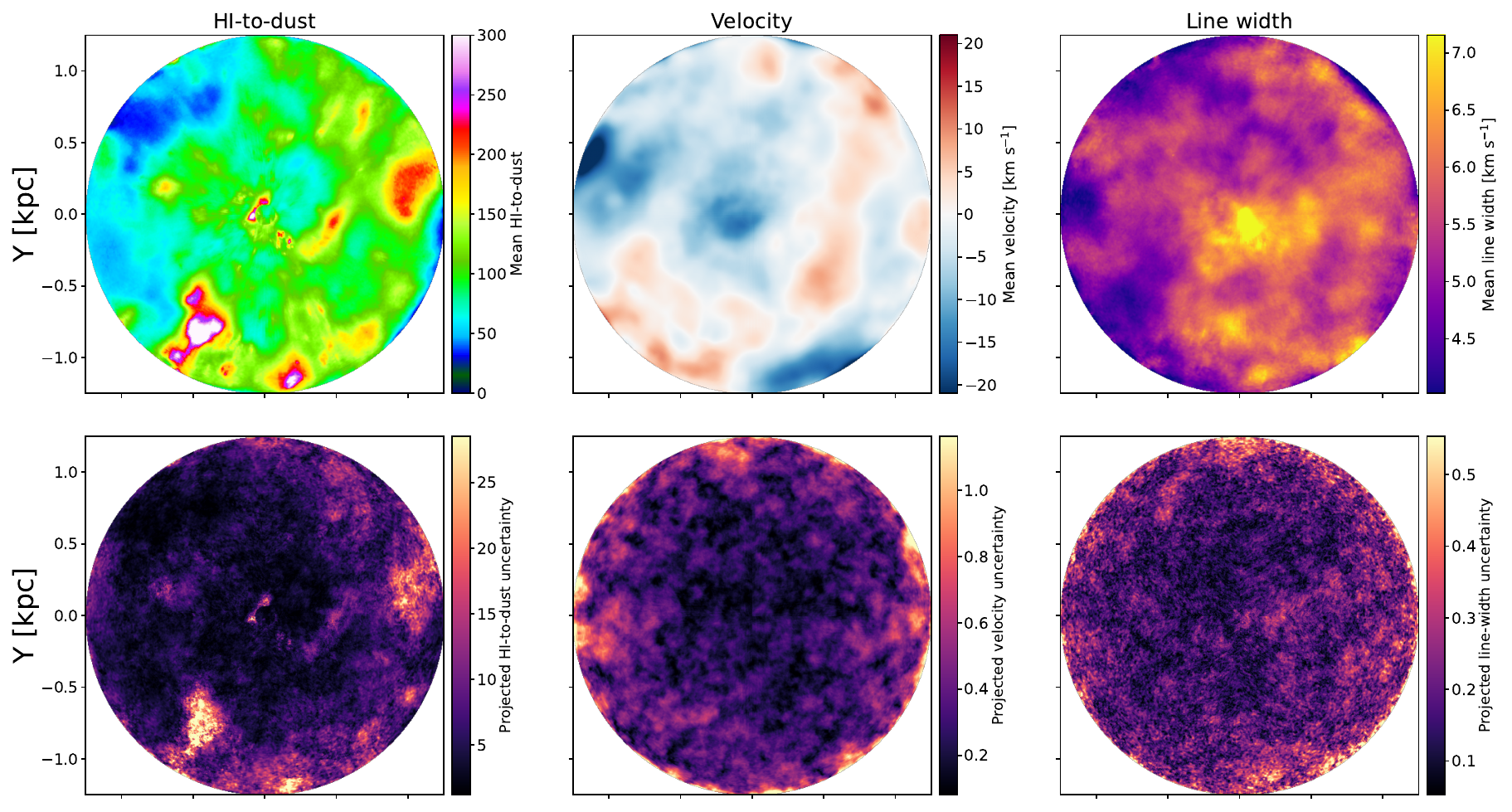}
    \caption{Visualisation of the three grids which make up our posterior mean reconstruction of the local \ion{H}{I}. The top row shows the mean value through the z-axis of the grids, with the bottom row showing the uncertainty in each image. The uncertainty maps are calculated by taking the standard deviation of projected map samples, rather than being a projection of the voxel-wise variance. The left column shows the \ion{H}{I}-to-dust ratios grid. The middle column shows the additive non-circular rotation velocity field. The right column shows the grid of effective line width. These views are all averaged through the $z$-axis in heliocentric Galactic coordinates, with the Galactic centre to the right.}
    \label{fig:threegrids}
\end{figure*}

\begin{figure*}
    \centering
    \includegraphics[width=1.0\textwidth]{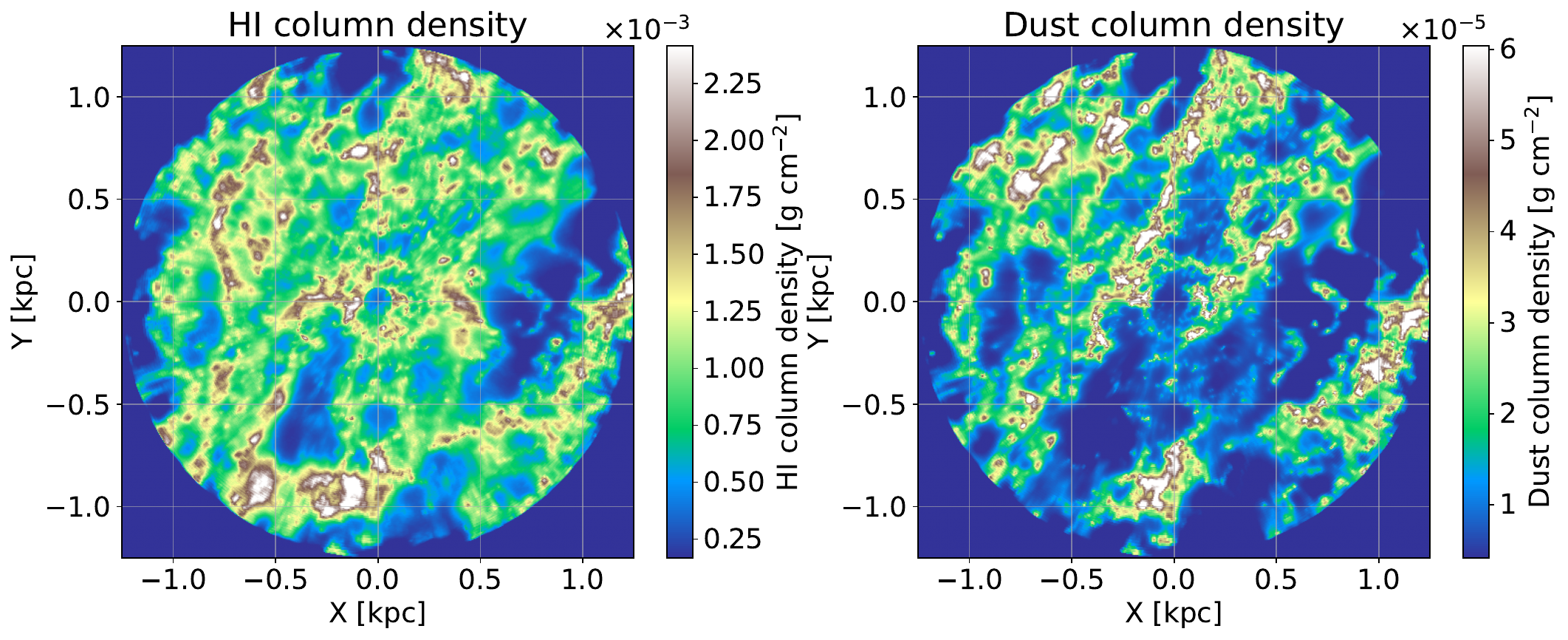}
    \caption{Left: Top-down view of linearly scaled \ion{H}{I} column density derived from our posterior mean \ion{H}{I} density grid. Axes are in heliocentric Galactic Cartesian coordinates, with the Galactic centre to the right. As this is a $Z$-integrated column density through a spherical reconstructed volume, pixels towards the edge of the mapped area are integrated through a smaller range of $Z$-values. This view of local \ion{H}{I} is one which is not traditionally observable. The central circular artefact is due to the unmapped volume at the centre of the map ($d<69~\rm pc$). Right: The top down view of the underlying \citet{edenhofer23} dust map. This map colour bar spans the same range divided by a factor of 40.}
    \label{fig:densitymap}
\end{figure*}

\begin{figure*}
    \centering
    \includegraphics[width=1.0\textwidth]{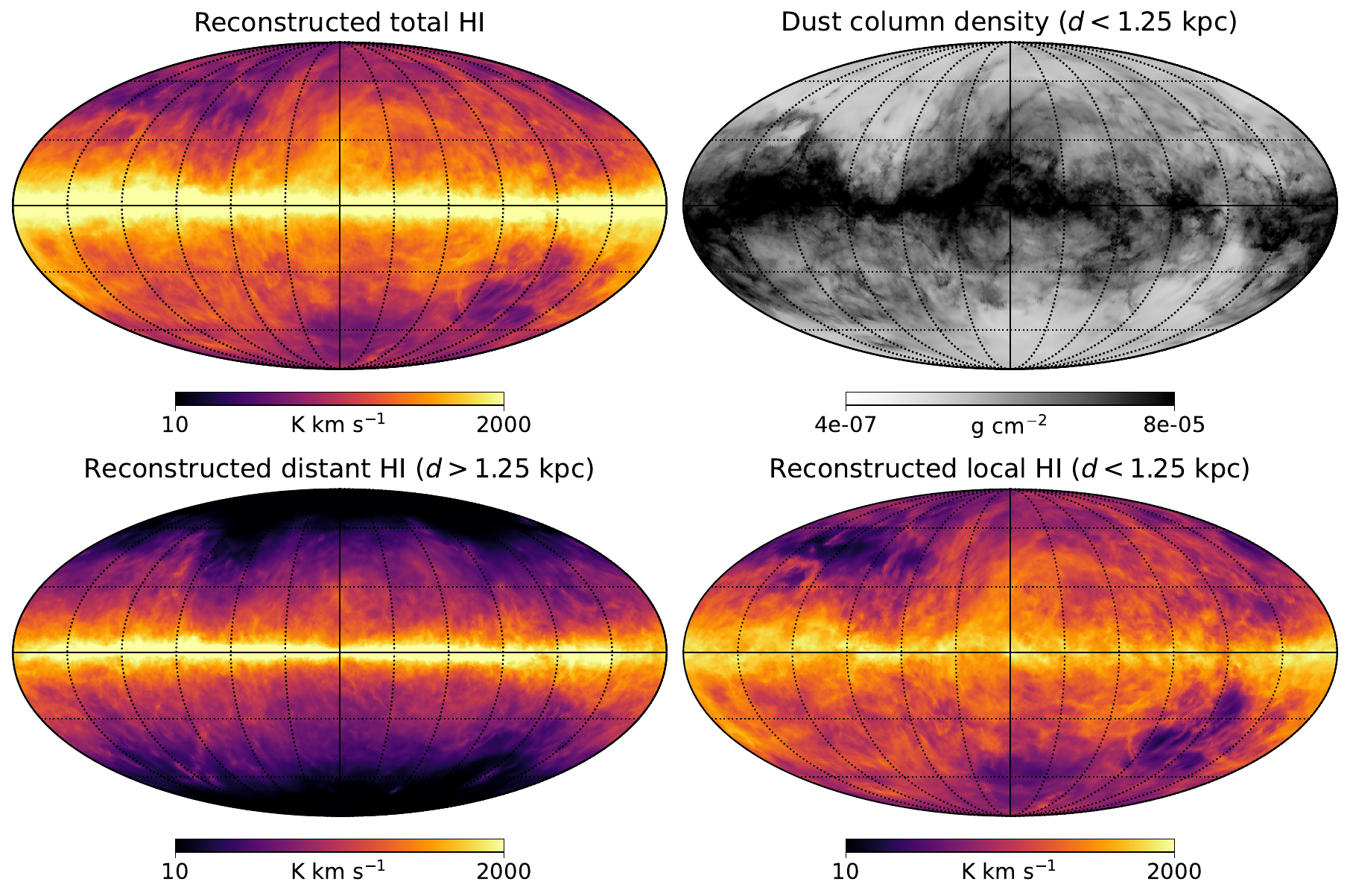}
    \caption{Decomposition between local and distant sky from our posterior mean grids, with all panels centred on the Galactic centre ($\ell=0^{\circ}$). Top left panel shows the total reconstructed sky, which matches the HI4PI dataset at $|V_{\rm LOS}| < 75 \rm ~km~s^{-1}$ within a reduced $\chi^{2}_{HI4PI}$ of 1.3. The top right panel shows the \citet{edenhofer23} dust map projected on the sky and converted to a total mass column density using our method. Bottom left panel shows our reconstructed distant sky (>1.25~kpc) and bottom right shows our reconstructed local sky (<1.25~kpc).}
    \label{fig:decomposition}
\end{figure*}

The benefit of using 3D correlated fields to model the ISM is that it informs the optimiser on which features are kinematically coherent which the regions around them, allowing us to separate out the distant and local skies. However this method will break down in certain regions of the local volume, specifically at or around higher latitude intermediate velocity clouds (IVCs). These structures represent small clouds which have been thrown high above the Galactic plane, and thus sit very anomalously in velocity space relative to their surroundings. Using the method as described above, the optimiser will place these structures in the distant sky. This is not because our model is unable to build such structures, but simply that they are hard to find in the parameter space. In order to address this, we adopt the following method for seeding the velocity structure from an initial guess based on morphological correlations between \ion{H}{I} and dust.

For each pixel on our output $N_{\rm side}=64$ HEALPix sky, we choose a $5^\circ$ aperture on which to evaluate local morphological correlations. The correlations are evaluated on an observational $N_{\rm side}=256$ sky grid, using a HEALPix disc aperture where all selected equal-area pixels are weighted equally. The radial grid is the same 196-bin broken-logarithmic grid used for the reconstruction, spanning $0.07$--$1.25~\mathrm{kpc}$, with inner bin widths of $3.36~\mathrm{pc}$ increasing logarithmically beyond $0.3~\mathrm{kpc}$ to $13.9~\mathrm{pc}$ at the outer edge. The HI4PI cube is sampled over $|v_{\rm LSR}|\leq75~\mathrm{km\,s^{-1}}$, corresponding to 117 velocity channels with spacing $1.2876~\mathrm{km\,s^{-1}}$. Before correlation, both the dust and \ion{H}{I} templates are transformed logarithmically and high-pass filtered by subtracting a $6^\circ$-FWHM smoothed version of the map, so that the comparison is between local morphology rather than large-scale gradients. For each aperture, radial bin, and velocity channel, we compute the standard unweighted Pearson correlation coefficient between the high-pass dust morphology and the high-pass \ion{H}{I} morphology, after mean subtracting and normalising each extracted image.

For each radial bin we first identify the velocity channel with the largest positive correlation. A seed is accepted only if this best correlation coefficient is at least 0.75. The assigned seed velocity is then taken as a correlation-weighted mean over the contiguous range of nearby velocity channels whose scores remain above 70 per cent of the peak value, capped at a maximum of five channels and weighted by the square of the correlation coefficient. This avoids assigning the seed velocity from a single noisy channel, and ensures we count contributions from neighbouring velocity channels which also correlate well with dust.

\section{Results}

\subsection{Synthetic Data Test}
\label{mocktest}

Here we display the results from our synthetic data test as described in section~\ref{mocktestdescribe}. Figure~\ref{fig:mockhistos} shows the comparison between ground truth values and reconstructed values. These comparisons are in the form of a histogram of truth versus reconstructed, where a perfect match falls on the 1:1 line. These histograms were computed volumetrically, with our spherical grid first interpolated onto a $256^{3}$ Cartesian grid, with each voxel contributing equally to the histogram. All four panels show a very good recovery of the ground truth synthetic data. The biggest departure from the ground truth is seen in the line-width panel, where the reconstruction overestimates the truth line width by a small systematic amount. We find through testing that different configurations of line-width structure can lead to similar looking results in data space, and because of this are cautious to draw conclusions from our reconstructed map of line widths.

This likely reflects the fact that the line-width field is more weakly constrained by the data than the density and velocity fields. Changes in density or centroid velocity produce direct changes in the predicted intensity or velocity-channel location of the emission, whereas moderate changes in effective line width can have only subtle effects after line-of-sight integration, and can be partly degenerate with unresolved velocity structure or optical-depth effects. The systematic overestimate in the synthetic test may therefore indicate that the recovered line-width field remains more prior-informed than the density and velocity fields.

Figure~\ref{fig:mockmain} shows maps of gas density weighted line width, gas density, \ion{H}{I}-to-dust column density ratio, and gas density weighted LOS velocity. Our ground truth maps are shown beside the reconstructed maps, with the ratio/difference between the two being shown on the right. We find excellent agreement throughout most of the grid, with a few differences in the gas density and velocity structure towards the Galactic centre (to the right on these plots). There is a noticeable region near the $-y$ extent of the gas density plots with overestimated gas densities. The velocity structure is also somewhat overestimated in this region of the map, by a few $\rm km~s^{-1}$ at the worst. We suspect this region is challenging as there is a lot of dust density near the edge of the local volume, suggesting there might be a lot of gas emission originating from just outside the 1.25~kpc volume. These regions just outside the volume are fairly kinematically coherent with the structures just inside the volume, causing confusion for the optimiser. However, these artefacts are limited, and the general match to the ground truth is very good.

Figure~\ref{fig:mockskies} also shows our ground truth versus reconstructed structures but on the plane of the sky in order to show the success of our near/far decomposition. Our local sky is a near perfect match, with some slight overestimations in the direction of the Galactic centre as seen in the face-on maps. The distant reconstructed-to-truth ratio at some high latitudes is approaching a factor of 10, but these are regions of near zero intensity in the truth distant sky. The reconstructed total sky is a near perfect match to the dataset the optimiser was given.

\subsection{Gas Densities}

The results section beyond here refers to our real data reconstruction, rather than our synthetic data test. Our method returns a set of 10 equally likely samples from the posterior. The posterior mean grids which make up our model have been visualised in figure~\ref{fig:threegrids}. This includes the \ion{H}{I}-to-dust grid, the radial velocity grid before adding circular Galactic rotation, and the grid of effective line width. 

A top down view of the mean density grid of these 10 samples is shown in figure~\ref{fig:densitymap}, alongside the posterior mean dust map from \citet{edenhofer23}. We find that the 3D structure of \ion{H}{I} is notably more diffuse than the 3D structure of dust, with fewer sharp features and a generally smoother structure. The colour bars in this side-by-side comparison span the same dynamic range, but scaled up/down by a factor of 40.

Our method reconstructs the HI4PI data to within a reduced value of $\chi^{2}_{HI4PI}=1.3$. Here $\chi^{2}$ is computed over all retained sky pixels and velocity channels by comparing the posterior samples to the HI4PI data, with residuals normalised by the assumed HI4PI data uncertainty, and divided by the number of data values. This 1.3 value is the mean $\chi^{2}$ through all 10 posterior samples. The maps of the decomposed local/distant \ion{H}{I} emission skies are shown in figure~\ref{fig:decomposition}. We find that the local sky ($d < \rm 1.25~kpc$) accounts for 50\% of the observed emission with $|V_{\rm LOS}| < 75 \rm ~km~s^{-1}$. This rises to 78\% when only considering emission above $|b| > 30^{\circ}$. The local emission sky contains much of the morphological structure of the dust map sky, and there are few to no structures which are present in both the distant and local skies, suggesting a clean decomposition.

\begin{figure}
    \centering
    \includegraphics[width=1.0\columnwidth]{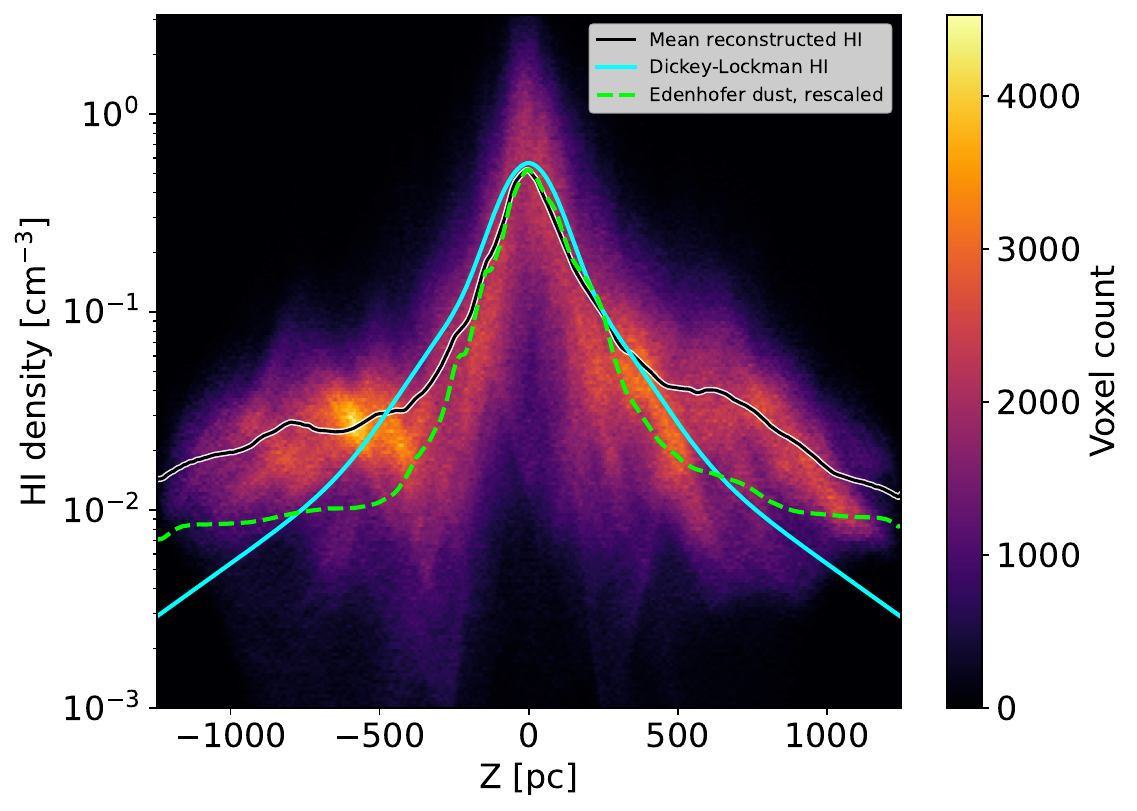}
    \caption{Comparison between our reconstructed vertical \ion{H}{I} profile and the expected \ion{H}{I} profile from \citet{dickey90}. Also shown is the profile extracted from the underlying \citet{edenhofer23} dust map, where the density has been scaled up to match the gas density in the midplane (a factor of 45). Note that the number of voxels included in this plot changes as a function of height due to the shape of our reconstructed volume, as the cross sectional area of our sphere decreases with $|z|$. Over-plotted is the 2D histogram of all voxels in the reconstructed posterior mean gas density map.}
    \label{fig:dickeylockman}
\end{figure}

We show the average \ion{H}{I} density as a function of $z$-height in figure~\ref{fig:dickeylockman}. This is shown alongside the fit to local \ion{H}{I} summarised by \citet{dickey90}. Compared to the \citet{dickey90} curve, we reconstructed slightly lower densities of \ion{H}{I} at $|z| < 500~\rm pc$, while seeing higher densities at higher altitudes. Due to the spherical shape of the reconstructed volume, the average densities at higher $z$-heights are less statistically significant (fewer voxels), and at the highest heights the average \ion{H}{I} density can be dominated by a single cloud/overdensity. However, we do find a significant systematically higher density of \ion{H}{I} at $|z| > 500~\rm pc$. 

\begin{figure}
    \centering
    \includegraphics[width=1.0\columnwidth]{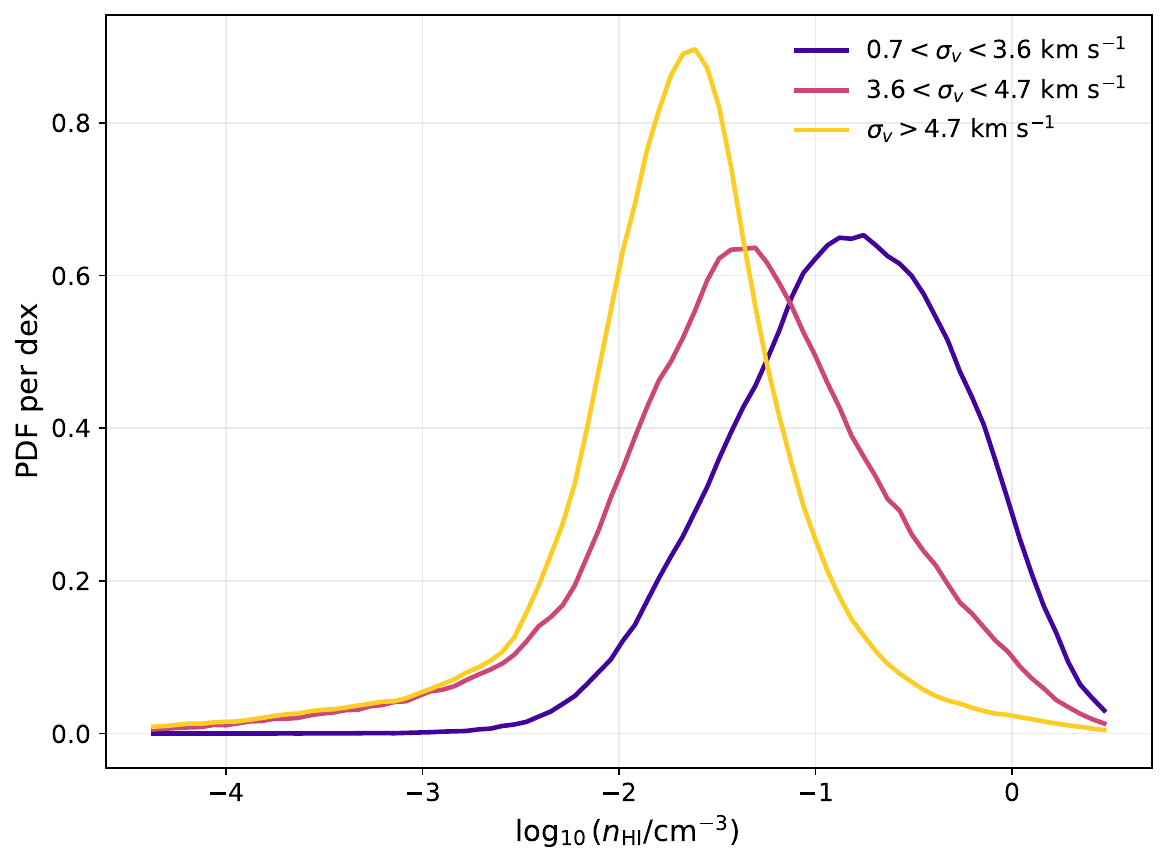}
    \caption{Probability distribution functions of \ion{H}{I} density, as a function of reconstructed line width. We find that voxels of higher line width tend to be occupied by lower \ion{H}{I} densities.}
    \label{densitypdfs}
\end{figure}

Figure~\ref{densitypdfs} shows the probability distribution functions for \ion{H}{I} density, separated into three bins by our reconstructed line width. These bins have been chosen to contain an equal number of Cartesian voxels in each one. We see a noticeable shift in mean density as we go to higher line widths.

Figure~\ref{fig:g2dhist} shows a 2D histogram of dust column densities versus \ion{H}{I}-to-dust column density ratios, where column densities have been evaluated through the $z$-axis giving a face on view. We include a horizontal line at \ion{H}{I}-to-dust = 100, representing the canonical gas-to-dust ratio of \citet{hensley23}, corrected to exclude the mass contributions of helium (140/1.4 = 100). This gives us the canonical prediction for total hydrogen-to-dust mass ratio. We find that our mean/median lines approximately trace the canonical value for dust column densities below around $10^{-5}~\rm g~cm^{-2}$, albeit with significant scatter. Above this density, the \ion{H}{I}-to-dust ratio drops off consistently, and at the highest dust column densities in our volume of around $10^{-4} \rm ~g~cm^{-2}$, the \ion{H}{I}-to-dust mass ratio is only around 20. 

We find a global \ion{H}{I}-to-dust ratio by dividing total mass of \ion{H}{I} by total mass of dust to be 50.5. For comparison, if we evaluate the total gas mass within our volume using the \citet{dickey90} profile, we get a global \ion{H}{I}-to-dust ratio of 58.0. We note that these global mass ratios are biased toward high density regions where molecular fractions are likely higher than global. We also find a volume weighted average \ion{H}{I}-to-dust mass ratio of 102.2.

We include a series of radially integrated shells of \ion{H}{I} column density in figures~\ref{fig:radialshells} and \ref{fig:radialshellsanti}.

\subsection{Gas Velocities}

We show the posterior mean structure of the \ion{H}{I} velocity in figure~\ref{fig:velocity}. This is a top down view of the map with Galactic centre to the right, showing the gas density weighted LOS velocity for each $X-Y$ position. This includes Galactic rotation. The expected quadrupole from Galactic rotation can clearly be seen, as well as other distinct regions of differing velocity to the underlying quadrupole. 

One feature seen is the expansion of the clouds which make up the local bubble within the nearest few hundred parsecs. Dense regions around the Sun are all seen to be moving away, as expected due to the expansion of the local bubble. We also include figure~\ref{localbubble}, which isolates the local bubble using the definition from \citet{oneill24}. We show the gas column density from all non-zero voxels in the \citet{oneill24} local bubble, and also the gas density weighted mean $V_{\rm LSR}$ along each line of sight. At low latitudes, velocities are overwhelmingly positive, consistent with the expectation for an expanding shell/bubble with the observer in the centre. 

Using the same \citet{oneill24} Local Bubble shell selection shown in Fig.~\ref{localbubble}, we can estimate the strength of this low-latitude LOS expansion signature. For each posterior sample, we compute the \ion{H}{I}-column-density-weighted mean $V_{\rm LSR}$ along each Local Bubble sightline. Restricting the resulting sky map to $|b|<10^\circ$, where the positive low-latitude shell dominates the \ion{H}{I} column, gives a column-density-weighted median $V_{\rm LSR}=+2.8~\mathrm{km\,s^{-1}}$ and a column-density-weighted mean $V_{\rm LSR}=+3.2~\mathrm{km\,s^{-1}}$. The column-density-weighted 16th--84th percentile range across low-latitude sightlines is $-0.7$ to $+7.6~\mathrm{km\,s^{-1}}$, and 79 per cent of the selected low-latitude Local Bubble \ion{H}{I} column lies in pixels with positive median velocity. These values should not be interpreted as a full three-dimensional shell-normal expansion velocity. The \citet{oneill24} shell is selected geometrically from the dust distribution, rather than kinematically, and can therefore include material that lies on the Local Bubble surface but is not part of the coherent expansion. This likely biases the simple LOS average below the $6.7^{+0.5}_{-0.4}~\mathrm{km\,s^{-1}}$ present-day expansion speed inferred by \citet{zucker22} from a dynamical model of the Local Bubble surface and young stellar-cluster motions.

A more surprising result is the switch to negative velocities at high latitude. We interpret these velocities cautiously, since the corresponding regions have lower reconstructed \ion{H}{I} columns than the low-latitude shell and are therefore more weakly constrained by the data. However, a high-latitude inward component would not be unprecedented. WHAM observations of diffuse H$\alpha$ show a net low-velocity infall toward the Galactic plane in both hemispheres \citep{haffner03,putman09}. The negative high-latitude \ion{H}{I} velocities in our reconstruction may therefore be related to a real fountain/chimney flow, although we do not claim a robust detection from the present \ion{H}{I} reconstruction alone.

Figure~\ref{fig:inferredrotation} shows the circularly averaged rotational component inferred from the reconstruction. To place the profile on an absolute velocity scale, we use a normalisation of $237\rm ~ km ~s^{-1}$ at the Solar radius. This is a different normalisation than is applied in the reconstruction of the map ($230~\rm km~s^{-1}$). As described previously, the underlying rotation curve in our reconstruction is only an approximation of the truth, with departures from this approximation learned by the non-circular velocity field. The $237\rm ~km~s^{-1}$ value comes from \citet{reid19} and is likely a more realistic estimate for the Solar circle, and is hence adopted for the post-hoc analysis and generation of figure~\ref{fig:inferredrotation}. We find that the inferred profile decreases with radius, and does so more steeply than would be expected from the \citet{reid19} rotation curve. The profile also exhibits substantial scatter in each Galactocentric radius bin, reflecting the fact that the reconstructed velocity field contains significant non-rotational velocities.

We also include a comparison to maser and young stellar cluster data of \citet{reid19} and \citet{hunt23} respectively. Figure~\ref{fig:vel_residuals} shows the observed versus predicted line of sight velocities. Error bars in the predicted velocities come from doing a Monte-Carlo sampling from each object's parallax uncertainty, as well as variance from the 10 different posterior samples. The observed velocity uncertainties are directly taken from \citet{reid19} and \citet{hunt23}. The right panel of this plot shows the residuals normalised to the combined uncertainty between observed and predicted velocity. We find a velocity data reduced $\chi^{2}$ value of 4.85 for the full dataset. This is the reduced chi-squared value computed from the residuals between the reconstructed and observed LOS velocities, with each residual normalised by the combined uncertainty of the observed tracer velocity and the reconstructed gas velocity. The reconstruction uncertainty includes both the effect of sampling over the quoted parallax uncertainty and the variance between the 10 posterior samples. Only 2\% of the objects are found outwith $5\sigma$ of their observed value, and 84\% of objects fall within $3\sigma$. 

For the set of masers we also include figure~\ref{masersvsvel}. This shows the volume weighted velocity in the reference frame of the \citet{reid19} A5 Galactic rotation curve, at $|z| < 150~\rm pc$. This is shown alongside the $|z| < 150~\rm pc$ maser positions, coloured by their observed velocities in the same reference frame.

The power spectrum of our velocity field is described by two parameters; the maximum length-scale above which correlations are suppressed, and the log-log slope of the power spectrum beyond this turnover length-scale. The posterior mean power spectrum for the velocity field has a turnover length-scale of 157~pc, and a log-log slope of -4.0 beyond this.

\section{Discussion}

Our resulting map of \ion{H}{I} emission is expectedly less structured than the underlying dust map which it is based on. Above certain densities, atomic hydrogen begins to form $\rm H_{2}$, meaning the densest dust clouds will not appear as dense in the \ion{H}{I} map. This can also be seen in the raw data, in which the dust map is more structured when projected at the same resolution as the HI4PI map. Unlike in \citet{soding25}, we do not reconstruct the CO density in an attempt to probe the molecular gas.

It is encouraging that our distant/local sky decomposition appears fairly distinct, with few structures shared between them. This might suggest that we have effectively separated structures rather than having them appear dimly in both skies. Inspection by eye also reveals that most structures in the dust map have been recovered in the local sky, with only a few isolated regions showing objects locally which do not appear in the dust sky, or vice versa. One notable exception to this is the presence of a cloud in the reconstructed local sky at very high negative latitude which does not appear in the dust map. This is likely part of the Magellanic stream \citep{putnam03}. This is a structure known to be far outside of our reconstructed volume, which the optimiser has erroneously placed locally. This is a challenging region for our method for two reasons. Firstly, the stream crosses zero velocity in this region, meaning it is kinematically coherent with local structures. Secondly, our remainder object was built with the shape of an exponential disc in latitude. The Magellanic stream is not part of the Milky Way's disc, and is therefore hard to explain in the distant sky at this intensity at this extreme latitude. 

The shift of the density PDF towards lower densities at higher line width is consistent with the idea that broader lines tend to be associated with more diffuse gas. We do not push this interpretation further here, since our effective line width includes both thermal broadening and unresolved motions, and is therefore not a direct measure of the underlying turbulence. We also see a broadening of the gas density PDF with decreasing line width, which is contrary to the trend expected from turbulence simulations. Nevertheless, the density/line-width trend suggests that the reconstructed map may capture some physically meaningful links between local gas density and thermal/kinematic state.

\subsection{Vertical Structure}

Our reconstructed vertical \ion{H}{I} profile is broadly consistent with the \citet{dickey90} benchmark, but shows slightly lower \ion{H}{I} densities at lower altitudes. While the \citet{dickey90} curve is not to be taken as a ground truth, we note that our reconstruction could underestimate midplane intensities if our local/distant decomposition were to assign some mid-plane emission to the remainder sky rather than to the local volume. From synthetic data testing, we found that any interchange between local and distant skies tends to happen at the boundary of our reconstructed volume at $1~ {\rm kpc} < d< 1.25~\rm kpc$, and rarely within the inner 1~kpc. This is likely because these edge regions are much less constrained by the data, as at our limited resolution the distant shells are harder to morphologically match between the dust map and gas data. The edge regions near the midplane also exist in a velocity regime which is more consistent with more distant emission. In our volume of interest, the highest velocities are found at the outer edge of the map, and these regions can be easily confused for gas which lives just outside the 1.25~kpc sphere and is of a similar velocity.

\begin{figure}
    \centering
    \includegraphics[width=1.0\columnwidth]{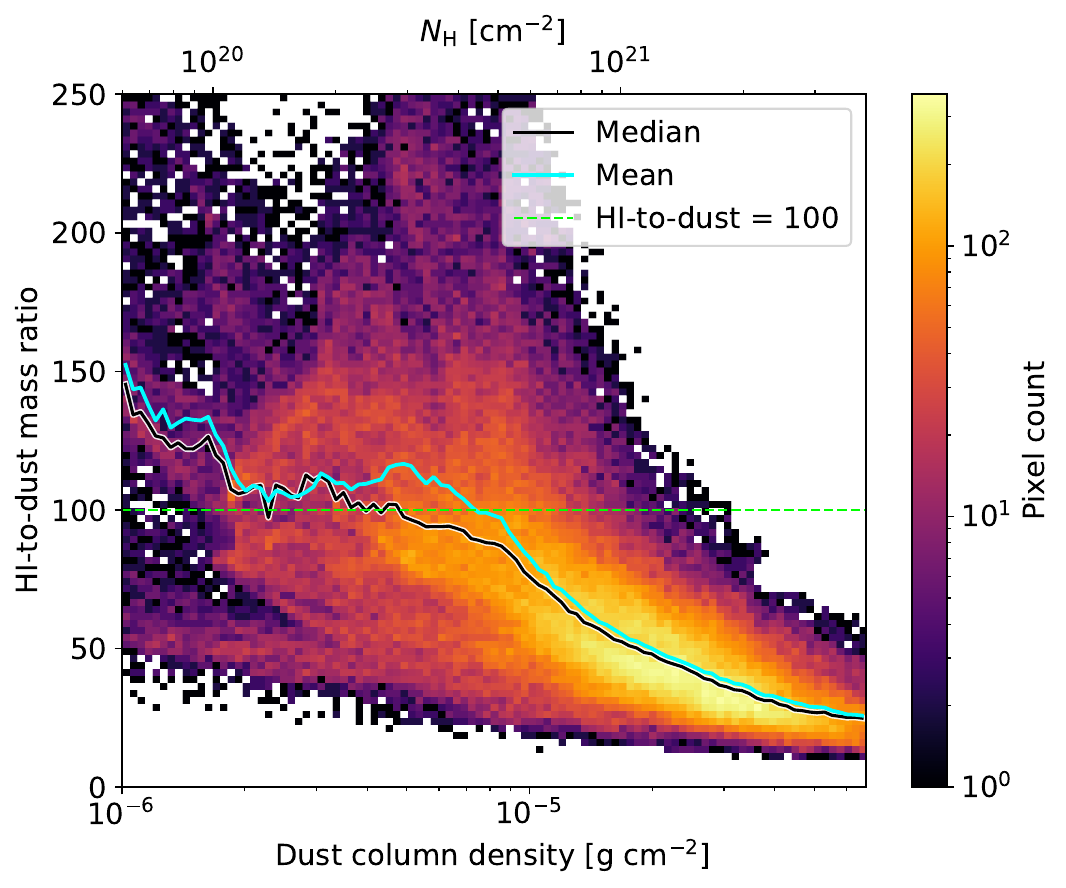}
    \caption{2D histogram showing the ratio of \ion{H}{I}-to-dust column density as a function of dust column density. Column densities were here taken as integrals through the $z$-axis, as if viewing the Galaxy from above. We show the median and mean \ion{H}{I}-to-dust ratio in each dust column density bin, as well as the expected hydrogen-to-dust ratio from \citet{hensley23}. Note that this value of 100 does not include mass contributions from helium, and has been adjusted from their total gas-to-dust ratio of 140. The secondary x-axis on the top of the plot assumes a hydrogen-to-dust ratio of 100.}
    \label{fig:g2dhist}
\end{figure}

\begin{figure*}
    \centering
    \includegraphics[width=1.0\textwidth]{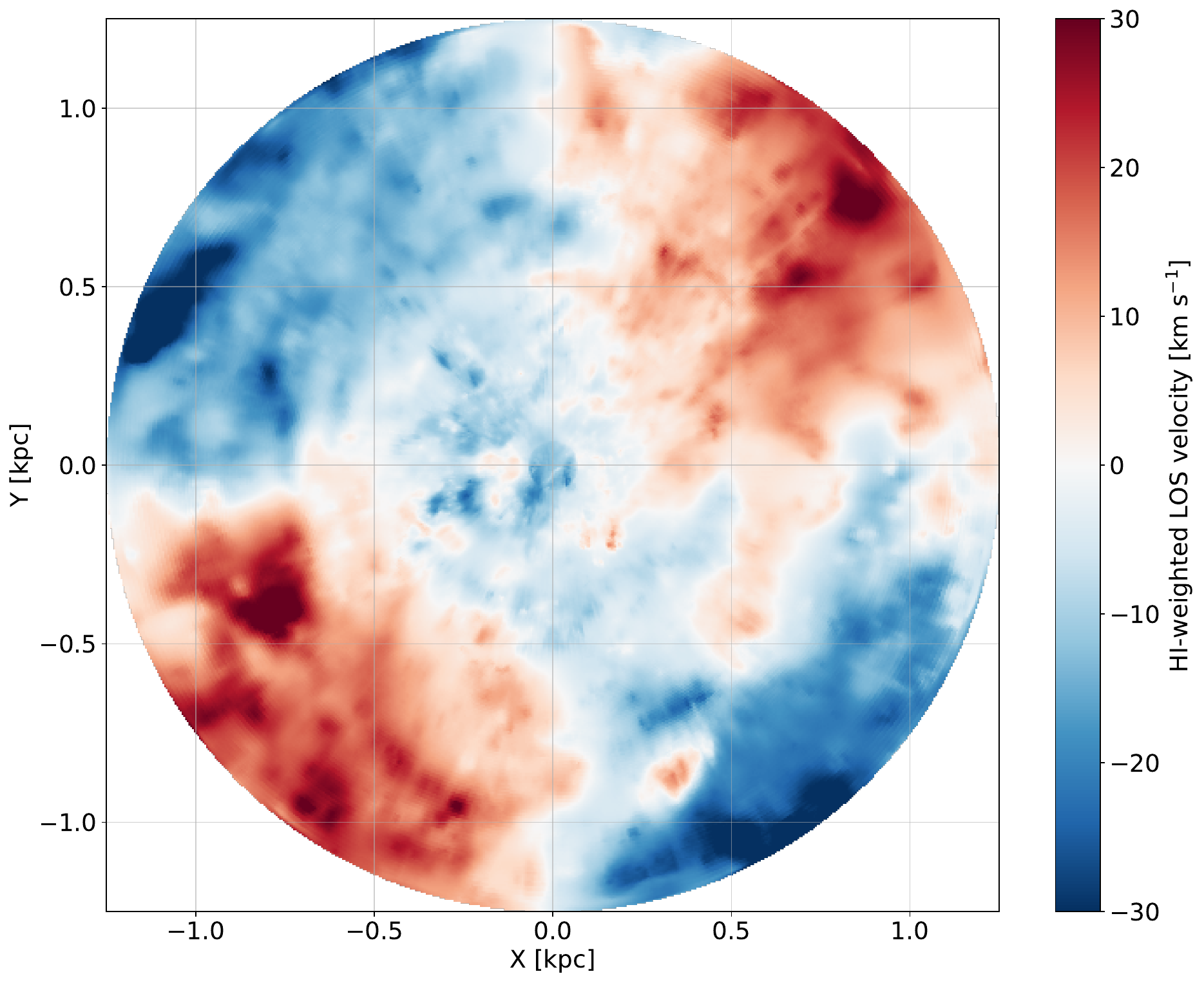}
    \caption{The same view of our reconstructed volume as in figure~\ref{fig:densitymap}, but showing the \ion{H}{I}-density weighted LOS velocity of the gas. The quadrupole expected from galactic rotation can clearly be seen, as well as a number of regions of high velocity relative to the underlying rotational velocity structure. Axes are in heliocentric Galactic Cartesian coordinates, with the Galactic centre to the right.}
    \label{fig:velocity}
\end{figure*}

We also recover higher number densities than the \citet{dickey90} benchmark at high altitude. We note that the negative-z side is near the region affected by contamination from the Magellanic stream, and that due to the spherical shape of our reconstructed volume, the number of voxels at extreme $z$ values is very few. This means just a small amount of contamination from outside the volume can throw the mean \ion{H}{I} density statistic to extreme values. Also worse voxel statistics at higher $|z|$ means that individual clouds can dominate the vertical density structure in figure~\ref{fig:dickeylockman}. While we certainly have contamination from the Magellanic stream, it appears that our reconstruction is consistently producing higher densities than \citet{dickey90} at high altitude. Much of the density in these regions is made up of isolated high altitude clouds, often at anomalous velocities relative to the surrounding medium (this includes Intermediate Velocity Clouds (IVCs)). In testing, we found that without the velocity seeding step described in section~\ref{seedsection}, these isolated clouds were not picked up by our reconstruction, and rather explained as distant emission. When these clouds are neglected, the Dickey-Lockman estimate is matched much closer. Only with the appropriate inclusion of these hard-to-capture clouds do we find the average high-altitude densities reported in figure~\ref{fig:dickeylockman}.

We emphasize that the \citet{dickey90} curve is used here only as a classical benchmark. It is not a direct measurement of the local \ion{H}{I} distribution within our reconstructed volume, but an empirical fit to the broader Galactic \ion{H}{I} layer at roughly the Galactocentric radius of the Sun. The profile is also vertically symmetric by construction and is intended as a compact descriptor. For these reasons, the aforementioned deviations from it are not necessarily evidence of a failure of our reconstructed 3D map.

\begin{figure*}
    \centering
    \includegraphics[width=1.0\textwidth]{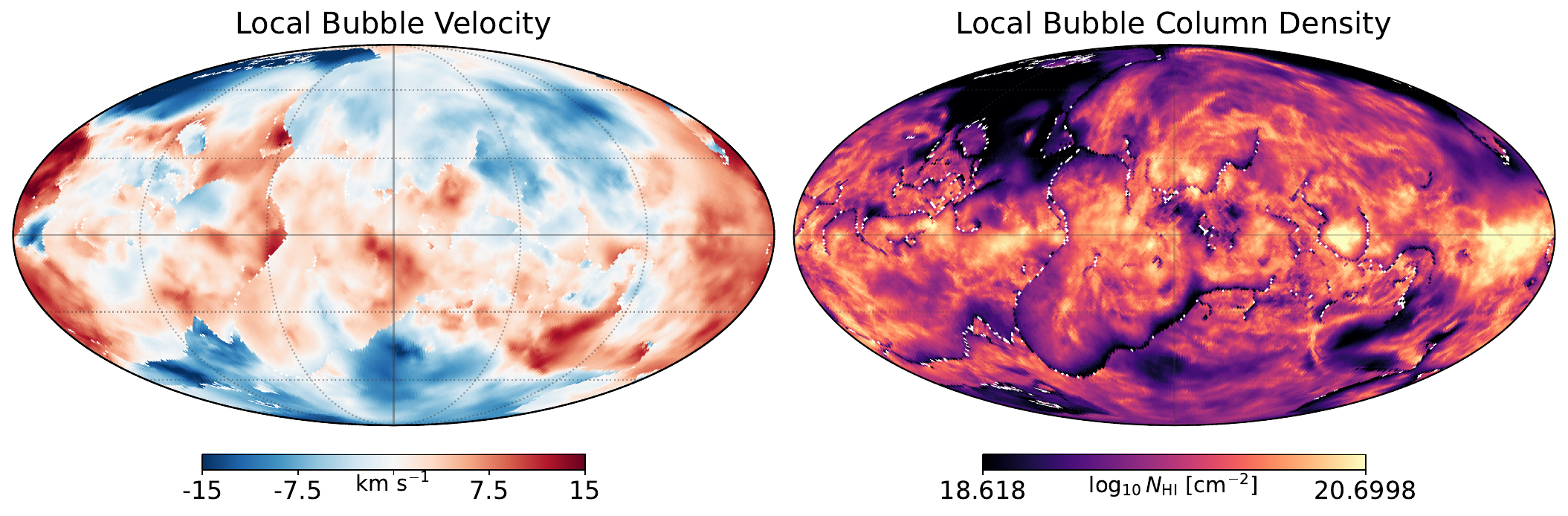}
    \caption{Left: the \ion{H}{I} density weighted mean $V_{\rm LSR}$ for all non-zero voxels from the local bubble definition of \citet{oneill24}. Colourbar is on a linear scale. Right: the total \ion{H}{I} column density from this same selected 3D region. Both maps are centred on the Galactic centre ($\ell=0^{\circ}$). Colourbar is on a logarithmic scale.}
    \label{localbubble}
\end{figure*}

\begin{figure}
    \centering
    \includegraphics[width=1.0\columnwidth]{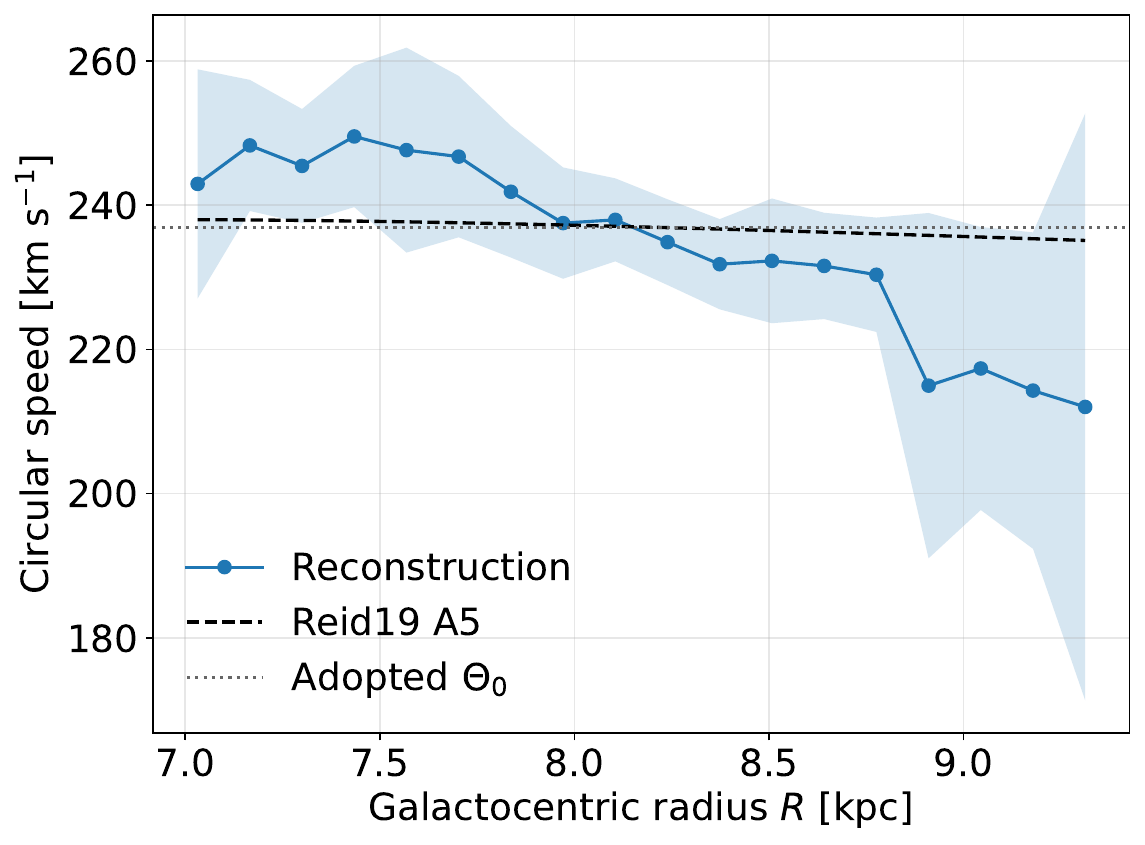}
    \caption{Radially averaged circular-rotation profile inferred from the reconstructed LOS velocity field, assuming an axisymmetric circular component and averaging in regions of constant of Galactocentric radius. To convert the reconstructed $V_{\rm LSR}$ field into an absolute circular speed, we adopt a local normalisation of $\Theta_0 = 237 \rm ~km~ s^{-1}$ at the Solar radius. Points show the annular mean circular component, while the shaded band indicates the \ion{H}{I} density weighted standard deviation of the inferred circular-speed values within each annulus. Over the limited range of radii sampled by the local reconstruction, the inferred circular component declines with radius much more steeply than the \citet{reid19} rotation curve.}
    \label{fig:inferredrotation}
\end{figure}

\begin{figure*}
    \centering
    \includegraphics[width=1.0\textwidth]{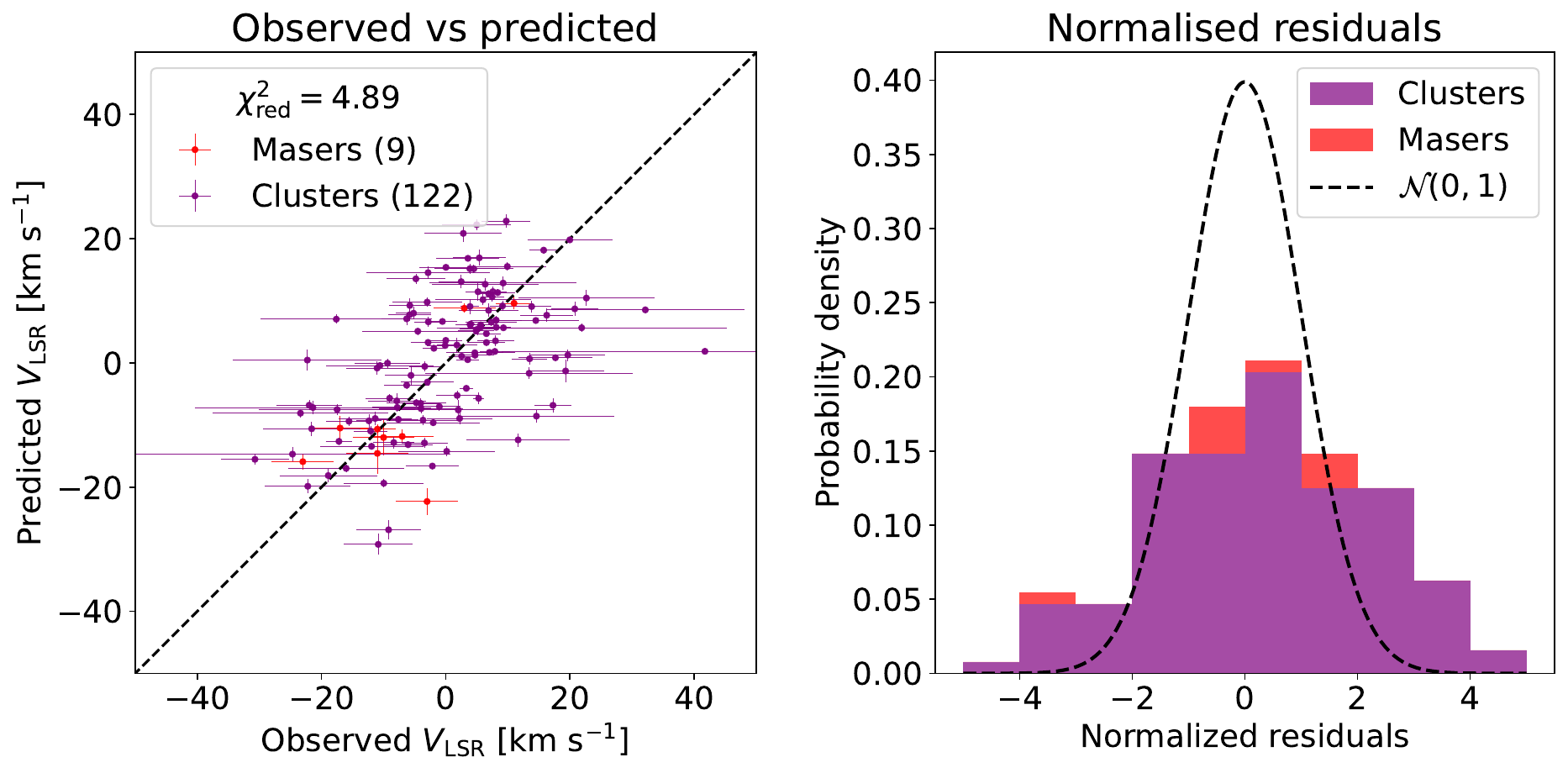}
    \caption{The left panel shows the LOS velocity of the dataset of masers and young stellar clusters (see section~\ref{vel_tracers}), versus the LOS velocity which is predicted by our velocity reconstruction. Errors in the $x$-axis are the measurement uncertainties on the observed data points, while errors in the $y$-axis come from the parallax uncertainties, and the variance between our distinct posterior samples. The right panel shows the residuals between observed and predicted velocities (predicted - observed), normalised to the joint uncertainty in both axes.}
    \label{fig:vel_residuals}
\end{figure*}

\begin{figure}
    \centering
    \includegraphics[width=1.0\columnwidth]{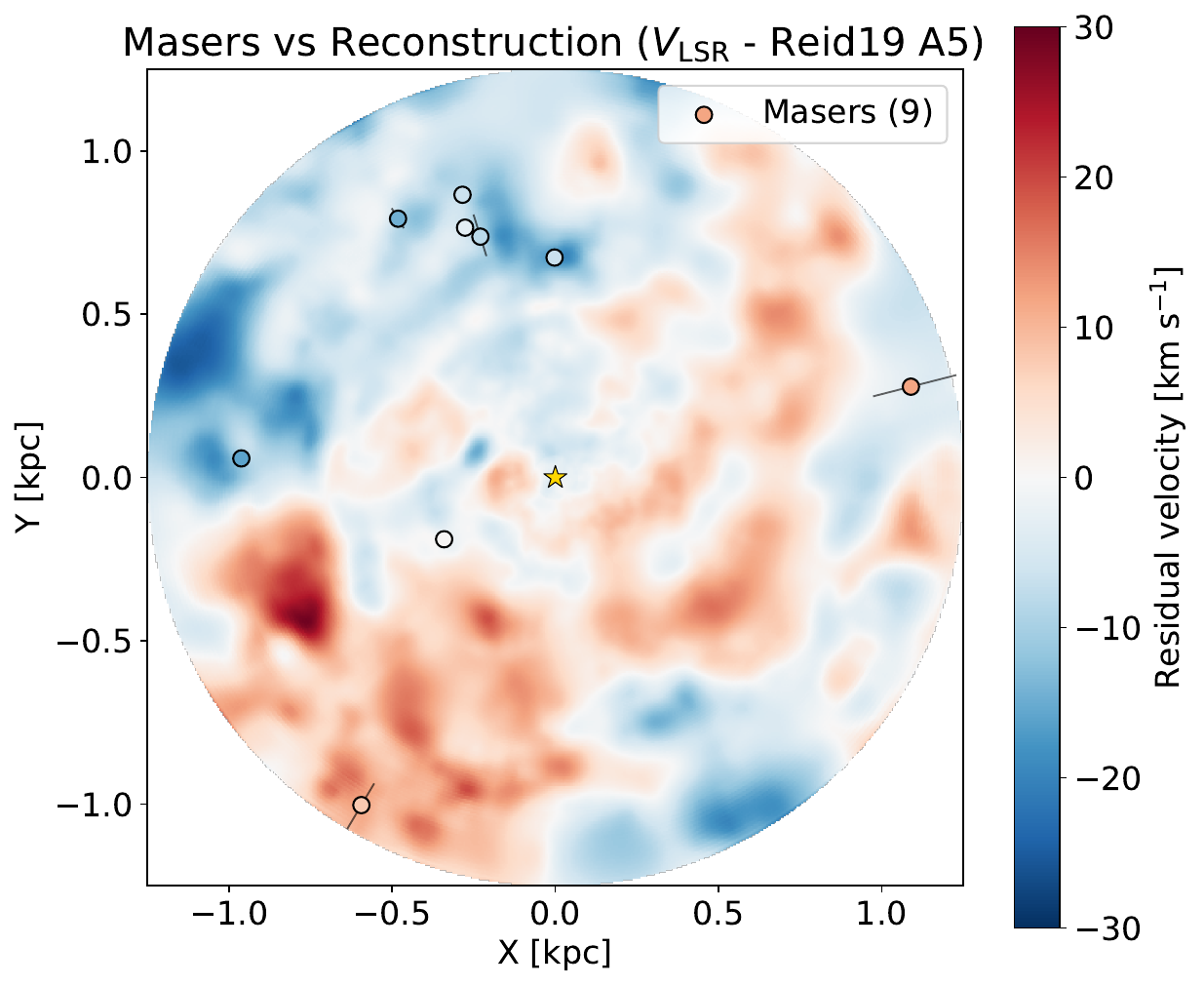}
    \caption{Top down view of our posterior mean reconstructed velocity field with the \citet{reid19} A5 Galactic rotation curve subtracted. This plot only contains cells of $|z| < 150~\rm pc$, and is a volume weighted average through the $z$-axis. Also included are the nine masers which fall below this $z$-ceiling, coloured by their observed velocities, also having subtracted the \citet{reid19} rotation curve. Axes are in heliocentric Galactic Cartesian coordinates, with the Galactic centre to the right.}
    \label{masersvsvel}
\end{figure}

\begin{figure*}
    \centering
    \includegraphics[width=1.0\textwidth]{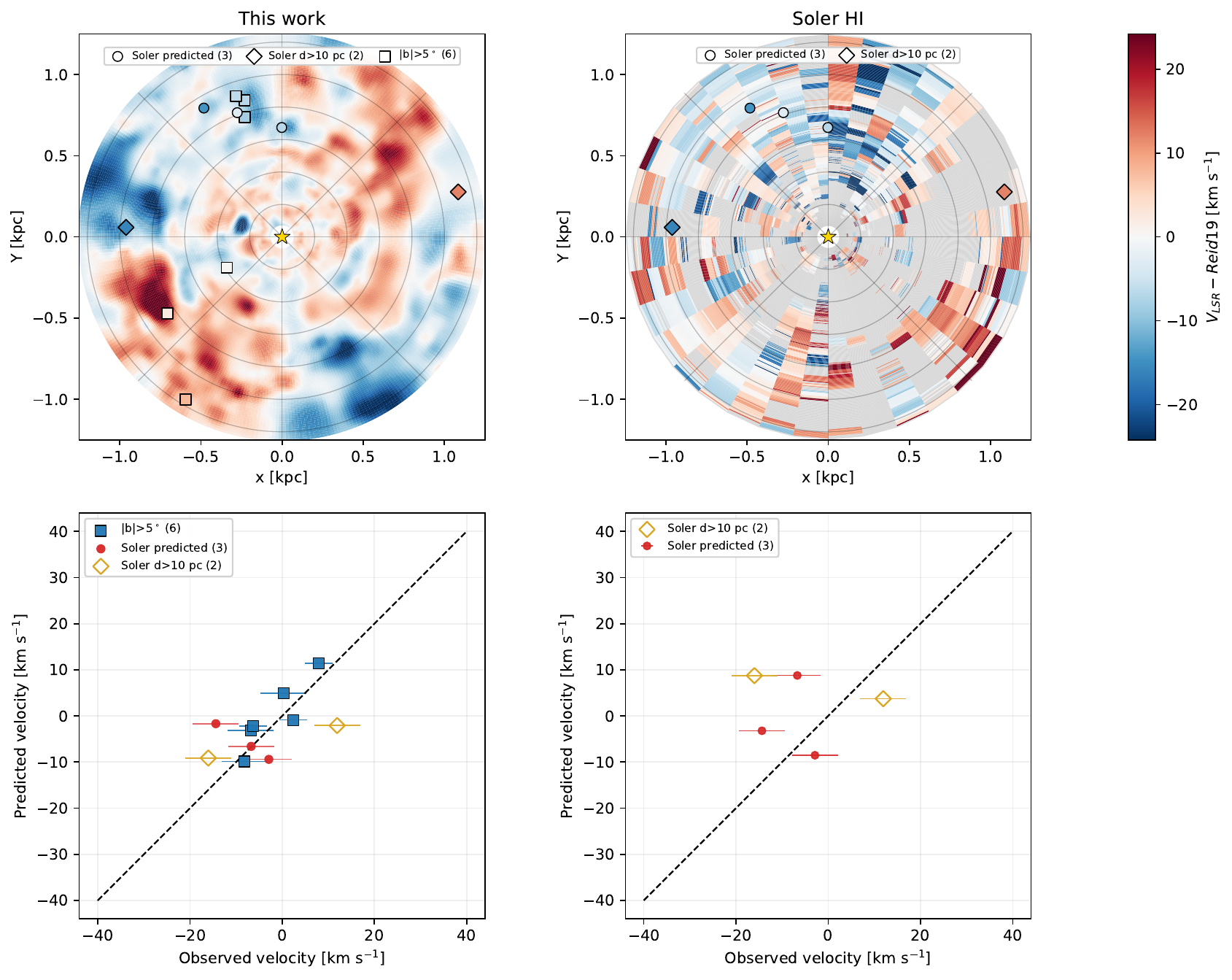}
    \caption{Top row shows the comparison in velocities from this work and \citet{soler25}. Our velocity structure has been volume-averaged through the z-axis, including only voxels at $|b| < 5^{\circ}$. Over-plotted are the six observed maser velocities within this volume and with $|b| < 5^{\circ}$. Two of these six are further then $10~\rm pc$ from the nearest valid Soler cell, and appear in diamond shape. The five square masers on the left plot are at $|b| > 5^{\circ}$ and are thus not predicted by \citet{soler25}, but are by our reconstruction. Rings are drawn every 200~pc in radius, and lines drawn every $45^{\circ}$ in Galactic longitude. The bottom row shows these same nine masers, comparing their observed velocities to those predicted by our reconstruction (left) and the Soler HOG method (right). Axes are in heliocentric Galactic Cartesian coordinates, with the Galactic centre to the right.}
    \label{fig:soler_grid_comparison}
\end{figure*}

\subsection{\ion{H}{I}-to-dust ratios}

While the global \ion{H}{I}-to-dust mass ratio of 50.5 might seem low compared to the modern total gas-to-dust estimate of 140 \citep{hensley23}, our ratio \emph{only} includes the neutral atomic hydrogen component of the ISM. A deficit relative to the total gas-to-dust ratio therefore reflects the gas in phases not reconstructed here, mostly molecular hydrogen in dense shielded regions and ionized hydrogen in ionized regions, as well as the helium contribution included in the total gas-to-dust ratio. If we assume a `true' gas-to-dust ratio of 140, and adjust for helium abundance by dividing by 1.4 gives us a total hydrogen-to-dust ratio of 100. Because of this the \ion{H}{I}-to-dust ratio conveniently reads directly as a \ion{H}{I}-to-total hydrogen mass percentage. See below:

\begin{equation}
    \frac{n_{\rm H_{I}}}{n_{\rm H_{\rm total}}} = \frac{n_{\rm H_{I}}/n_{\rm dust}}{n_{\rm H_{\rm total}}/n_{\rm dust}}
\end{equation}

where $n_{H_{\rm total}}/n_{dust} = 100$ \citep{hensley23}. Therefore our mass weighted average \ion{H}{I}-to-dust ratio of 50.5 suggests that 50.5\% of the hydrogen by mass is in the neutral atomic phase. \citet{kalberla09} quotes that galaxy-wide, atomic hydrogen is around 64\% of the mass of all hydrogen in the Milky Way. This way of thinking also lends itself to interpreting figure~\ref{fig:g2dhist}, where the mean/median drop-off at high dust density can be read as the decrease in \ion{H}{I}/$\rm H_{2}$ ratio as density rises, getting as low as 20\% in the densest columns of our reconstruction ($\approx 7\times10^{-5} ~\rm g ~cm^{-2}$).

From figure~\ref{fig:g2dhist} we can also read the dust column density at which \ion{H}{I} begins to transition to $\rm H_{2}$ at around $10^{-5}~\rm g~cm^{-2}$. \citet{krumholz09} finds through modelling the photodissociation of $\rm H_{2}$ that a gas column density of $10\rm ~M_{\odot}~pc^{-2}$is required to shield the formation of molecular hydrogen. Converting this to CGS units and dividing by the \citet{hensley23} gas-to-dust ratio of 140 gives a dust column density of $1.5~\times 10^{-5}~\rm g~cm^{-2}$, in very good agreement with our reconstruction. We can also compare to the results of \citet{krumholz09} at our highest dust column density of $\approx 7\times10^{-5} ~\rm g ~cm^{-2}$. We convert this to a total gas column density of around $50 \rm ~M_{\odot}~pc^{-2}$, and then extract the molecular hydrogen mass fraction from figure~2 of \citet{krumholz09}. At this density they predict a mass fraction of $\rm H_{2}$ of around 0.72. This is also in very good agreement with our lowest \ion{H}{I}-to-dust ratios, which imply around an 80/20 split of molecular to atomic hydrogen at this dust column density. 

Observational efforts also yield similar numbers, with \citet{shull21} finding the transition to $f_{\rm H_{2}}>0.1$ at approximately $N_{\rm H} > 10^{21} ~ \rm cm^{-2}$. It appears that we have successfully recovered the expected transition from atomic to molecular hydrogen as column density increases. We also highlight that this was done by evaluating the column densities over a spatial axis which is not traditionally accessible by observations (summing through the recovered $z$-axis), and is further validation of the success of our 3D reconstruction of \ion{H}{I} densities.

There are some likely effects of our assumed constant \ion{H}{I} spin temperature of 200~K. In very dense, cold regions this is likely too high a value, giving too low an optical depth between emitting gas and the observer. This disallows the optimiser from recovering the densest regions to their true values. This in some sense can be considered like a saturation effect, whereby the densest clouds are reconstructed at lower densities than in reality. The opposite is likely true for very diffuse media, but these regions are mostly optically thin regardless of the assumed $T_{\rm spin}$.

We also note that the densities reported throughout this work are volume-averaged \ion{H}{I} densities at the effective resolution of the reconstruction. They should therefore not be compared directly to sub-parsec CNM densities, which can reach $\sim10$--$50~\mathrm{cm^{-3}}$. A voxel may contain a mixture of phases and sub-grid structure, so our density field represents the average atomic hydrogen content on the resolved spatial scale.

Other recent 3D dust and extinction maps could also be considered for this purpose. For example, \citet{vergely22} provide larger Cartesian volumes around the Sun, and \citet{wang25} also extends all-sky extinction mapping to substantially larger distances. These maps are valuable for large-scale Galactic applications, but the \citet{edenhofer23} map is particularly well matched to the present reconstruction:  in particular for it's high spatial resolution within the local region we seek to reconstruct here. The \citet{edenhofer23} map was proved to work well for this type of local application in the 3D H$\alpha$ modelling of \citet{mccallum25}, where it reproduced much of the observed local WHAM H$\alpha$ morphology. We therefore adopt \citet{edenhofer23} as the dust-density anchor for this work, while noting that larger-volume dust maps will be valuable for future extensions beyond the local volume.

Some features of the \citet{edenhofer23} dust map will also imprint themselves on our recovered gas-to-dust ratios. The Gaia derived extinctions are known to saturate when extinction begins to completely obscure the stars used to determine the values, and thus the highest dust densities are not truly recovered in the map. This saturation effect will appear as elevated gas-to-dust ratios in our reconstruction. The \citet{edenhofer23} map was also constructed using a statistically homogenous random field, with no disc structure imprinted as a prior. Because of this, the existence of the disc drives the random field higher in regions poorly constrained by data (i.e. at high altitudes). The extinction reconstructed at high-$z$ is thus likely too high, which will in turn drive our reconstructed high altitude gas-to-dust ratios lower.

Ionized hydrogen will also contribute to the missing gas budget in some environments, especially around hot stars and superbubbles. However, the transition from neutral to ionized gas is not expected to occur in any one specific density regime as with the molecular gas, but will be dependent on other more complex factors such as proximity to sources of ionizing photons, or the presence of shocks/SNe.

\subsection{Reconstructed Velocities}

Our velocity structure yields a velocity data reduced $\chi^{2}$ value of 4.85 for the combined population of young stellar clusters and masers. As discussed earlier, the 11 maser sources are likely a more reliable measure of the underlying gas velocity as the masing molecules are actually part of the gas. When using only these 11 sources we match the dataset with a reduced $\chi^{2} = 2.4$, with this number being dominated by only two poorly explained maser velocities, both of which exists in regions of the map with little-to-no \ion{H}{I} emission. The worse match for the young stellar clusters is likely due to either 1) overconfident error bars on the young stellar cluster LOS velocities, or 2) the fact that they are already decoupled from the gas velocities, even at ages <10~Myr. We find that if we introduce a systematic uncertainty to each data point of $3 \rm ~km~s^{-1}$, it aligns our cluster match quality to those of the masers as measured by the reduced $\chi^{2}$ value.

This steeper decline in rotational velocity than \citet{reid19} should be interpreted cautiously. Our reconstruction only probes a relatively narrow range of Galactocentric radii around the Solar neighbourhood, and the statistics worsen toward the edges of the reconstructed volume where the averages are based on fewer voxels.

While our reconstruction is found to exhibit a slightly steeper power spectrum than is expected from a pure \citet{kolmogorov41} cascade, this difference should not be used to exclude Kolmogorov-like turbulence. The observed spectral index might be modified from the gas intrinsic 3D velocity spectral index by various effects, for example that we are only probing the radial velocity rather then the full 3D velocity field. The length-scale turnover parameter also can also influence the effective slope of the kernel. If this parameter moves to smaller values the slope of the power spectrum is effectively shallowed, and the log-log slope can steepen to account for this. We also highlight that our velocity field does not probe only the field associated with the expected turbulent cascade in the Galactic midplane, but also picks out smaller individual clouds at high altitude. This introduces small-scale power to the velocity field, and makes it even harder to learn about the size scales of turbulence from the power spectrum alone. Finally, we note that \citet{kolmogorov41}-like turbulence was originally derived for incompressible fluids, which is likely a poor assumption for most phases of the ISM.

\subsection{Comparison to \citet{soler25}}

The most direct comparison for our reconstructed velocity field is the kinetic-tomography map of \citet{soler25}, who used their astroHOG method to associate distance-resolved dust structures with velocity-resolved \ion{H}{I} emission. The HOG reconstruction assigns a velocity to each distance channel in each \(10^\circ\times10^\circ\) Galactic-plane tile, while our method infers a continuous 3D velocity field through a forward model. The comparison must be made on similar spatial scales, we therefore average our reconstruction onto the \citet{soler25} grid before comparing the two velocity fields.

Figure~\ref{fig:soler_grid_comparison} shows the comparison of recovered velocity fields after subtracting the motions due to Galactic rotation, for which we used the \citet{reid19} A5 Galactic rotation model. In the frame of the local standard of rest, the two maps show good agreement, both recovering the bulk motions of the gas due to Galactic rotation, however after subtracting Galactic rotation more differences are apparent. The main qualitative difference is the spatial continuity of the inferred velocity field. The HOG map shows sharper changes between neighbouring distance channels and, in some regions, reversals from one cell to the next. This behaviour is expected for a method that assigns the velocity from the channel with statistically the best correlation: if different structures dominate the projected morphology in adjacent distance channels, the selected velocity can change abruptly. Our reconstruction instead favours a continuous velocity field unless sharper reversals are required by the data. A visual comparison with the simulated LOS-velocity maps in \citet{abboudeh26}, (see their fig.~B.1), suggests that the underlying simulated flow is somewhat smoother than the cell-to-cell structure of the HOG reconstruction. This might suggest the sharpest apparent reversals in the HOG map may overstate the abruptness of the physical velocity field, while our reconstruction is likely the smoothest possible field which also is compatible with the data.

Quantitatively, we compare the two reconstructions using the rotation-subtracted velocity field on the \citet{soler25} grid. The resulting velocity dispersions and energy densities are summarised in Table~\ref{tab:soler_comparison}. Since these quantities depend on the adopted spatial grid, velocity definition, and averaging scheme, they should be interpreted as approximate comparisons rather than exact like-for-like measurements.

\begin{table*}
\centering
\caption{Summary of velocity-dispersion and energy-density estimates compared to \citet{soler25}. All quantities refer to the non-circular LOS component unless otherwise stated. Native-grid refers to our Healpix-logR grid structure.}
\label{tab:soler_comparison}
\begin{tabular}{lcc}
\hline\hline
Quantity & This work & \citet{soler25} \\
\hline
Velocity dispersion on Soler grid & $6.0~{\rm km~s^{-1}}$ & $10.8~{\rm km~s^{-1}}$ \\
Mean kinetic energy density on Soler grid & $0.15~{\rm eV~cm^{-3}}$ & $0.19~{\rm eV~cm^{-3}}$ \\
Native-grid kinetic energy density & $0.18~{\rm eV~cm^{-3}}$ & -- \\
Line-width-derived energy density & $0.046~{\rm eV~cm^{-3}}$ & -- \\
Total kinetic plus line-width energy density & $0.29~{\rm eV~cm^{-3}}$ & -- \\
\hline
\end{tabular}
\end{table*}

The dispersion of the non-circular velocities is lower in our Soler-gridded reconstruction than in the HOG reconstruction, while the corresponding kinetic energy densities are similar. On our native grid, using volume weighting, we obtain a comparable kinetic energy density. We also estimate an additional contribution from the reconstructed line-width field, although this value should be interpreted cautiously because the synthetic data tests show that line-width structure is less consistently recovered than density or velocity structure.

The side by side comparison also highlights some common features and clear differences between the two reconstructions. Both maps recover a clear reversal toward \(80^\circ<\ell<90^\circ\) at \(d\simeq 800~\mathrm{pc}\), where \citet{soler25} identify diverging \ion{H}{I} motions near the North America nebula/W80 region. Both show coherent structure around the Vela region near \(\ell\simeq270^\circ\) and \(d\simeq800\)--\(1100~\mathrm{pc}\). Some differences appear in the fourth quadrant, especially around \(320^\circ<\ell<340^\circ\), where \citet{soler25} associate high velocity motions with the Ara OB1 region. In regions where the HOG reconstruction has no assigned velocity because the dust--gas morphological correlation is weak, our correlated-field prior still returns a continuous velocity estimate. These filled-in regions should be interpreted with care, since the reconstruction there is expected to be more prior-driven.

\begin{figure}
    \centering
    \includegraphics[width=1.0\columnwidth]{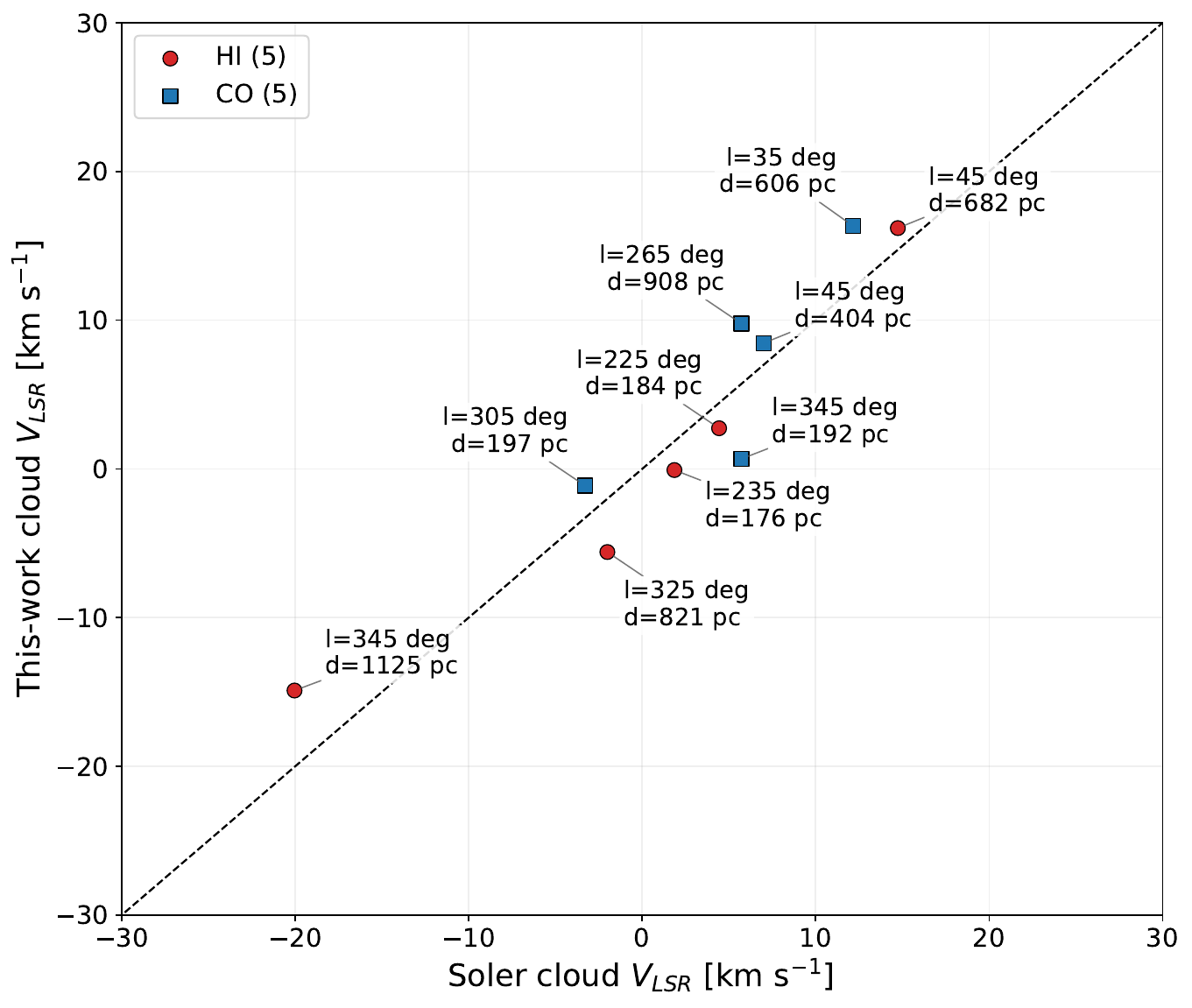}
    \caption{Comparison of ten $V_{\rm LSR}$ values as detected by the Soler HOG method, and as predicted by our reconstruction. These ten regions are the five most confident results of the HOG method using \ion{H}{I} data, and the five most confident results from the CO data. A very strong correlation is seen between the gold standard HOG results and our reconstruction.}
    \label{fig:soler_confident_clouds}
\end{figure}

 A second comparison is provided by the highest-confidence HOG cloud velocities identified by \citet{soler25}. This comparison is useful because HOG assigns velocities statistically to dust structures based on morphological alignment in PPV space, rather than solving a fully regularised 3D inversion problem. These clouds are therefore a good test case for our method, since they correspond to dust--gas associations that are visually and statistically best supported. For each cloud, we average our reconstructed velocity over the same sky and distance range used for the HOG calculation, and compare this predicted velocity to the HOG velocity in Fig.~\ref{fig:soler_confident_clouds}. The resulting agreement shows that the two methods identify broadly consistent velocities for the clearest structures in the local Galactic plane. This demonstrates that the smoother appearance of our velocity field is not simply caused by washing out the velocity signal: where the HOG confidence is strongest, our forward-model reconstruction recovers very similar cloud velocities.

A further test is provided by independent velocity tracers in the form of the \citet{reid19} masers. Figure~\ref{fig:soler_grid_comparison} also compares the predicted velocities from both reconstructions to the masers of \citet{reid19} and to the calibrator clouds used in this work. These tracers are not the necessarily the target of the HOG reconstruction, which measures prevalent velocities over larger \(10^\circ\times10^\circ\) regions, but they provide an independent check on the local gas kinematics.

\subsection{Intermediate Velocity Clouds (IVCs)}

\begin{figure*}
    \centering
    \includegraphics[width=1.0\textwidth]{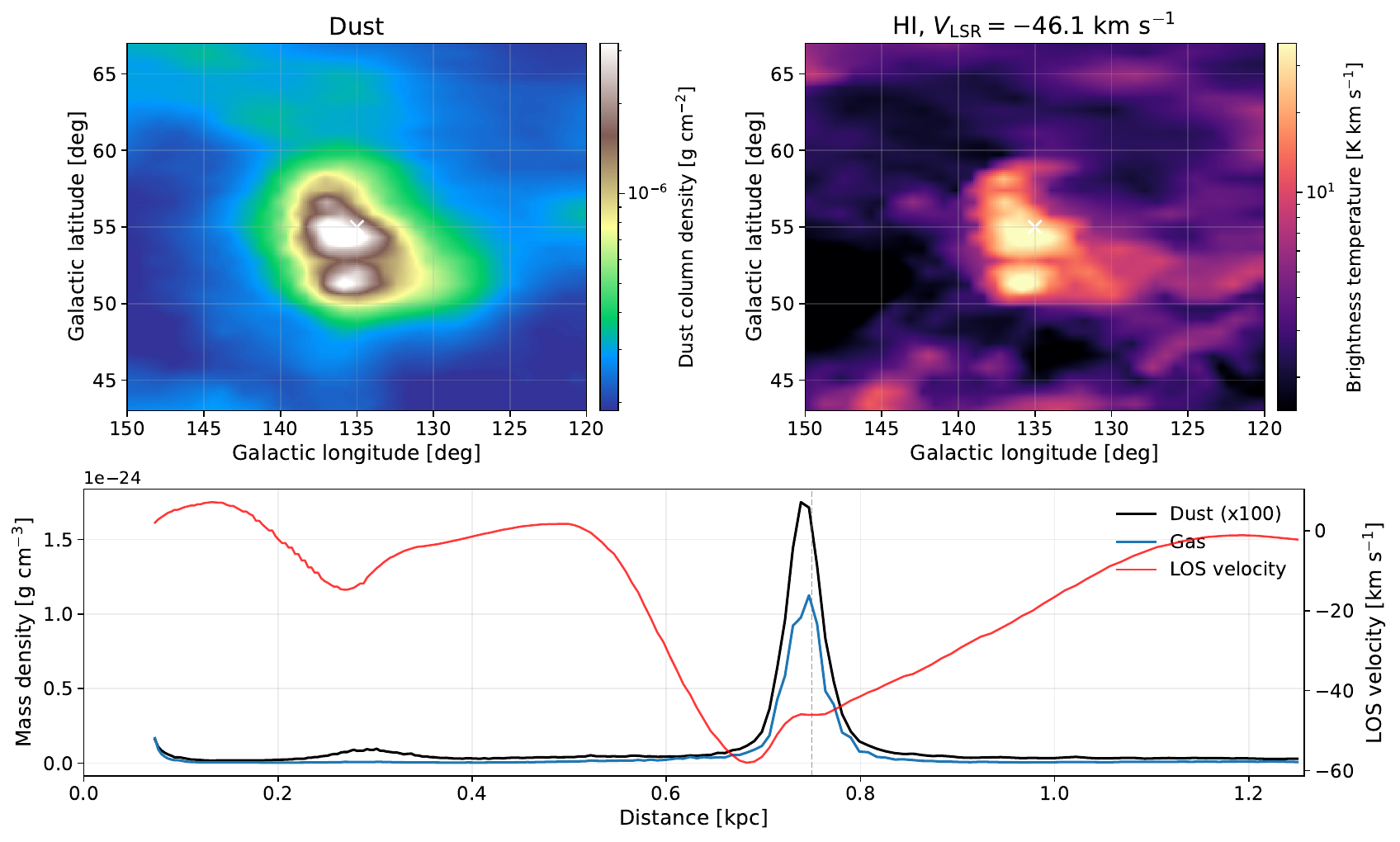}
    \caption{IVC 135 as seen in dust (top left) and \ion{H}{I} data in the velocity channel at $V_{\rm LSR} = -46.2~\rm km~s^{-1}$. Bottom panel shows the full line of sight profile of the reconstructed velocity, reconstructed gas density, and \citet{edenhofer23} derived dust density.}
    \label{ivc135}
\end{figure*}

In section~\ref{seedsection} we describe the introduction of a pre-processing step in order to make the identification of kinematically incoherent, high-latitude structures easier for our reconstruction to place locally. We use the known IVC at $\ell=135^{\circ}, ~b=55^{\circ}$ (IVC 135) as an example. This cloud was earlier thought to lie at a distance of $300$ -- $400~\rm pc$ \citep{benjamin96,hernandez13}, but morphological agreement between the dust map and \ion{H}{I} data places this cloud closer to $750~\rm pc$ in distance. Figure \ref{ivc135} shows the quality of the match in morphology between the dust map at this distance, and the \ion{H}{I} data at $V_{\rm LSR} = -46.2 \rm ~km~s^{-1}$. The bottom panel of this figure then shows the results of our reconstruction along the line of sight through the centre of IVC 135. It is seen that we match the expected velocity of the cloud at the expected distance. This figure also illustrates another feature of our reconstruction, which is the tendency for extreme velocities to hide in regions of very low \ion{H}{I} density. Both in front of and behind IVC 135, the extreme velocity associated with this cloud propagates into the low density space with little consequence for the observed data. This highlights that gas velocity reliability is likely linked strongly to the reconstructed \ion{H}{I} density.

\subsection{Future Possibilities}

This data product may open new avenues for improvement in other areas. The 3D map of ionized gas simulated in \citet{mccallum25} assumed a constant dust-to-gas ratio. Our reconstruction does not directly constrain the ionized component, but it does show that the \ion{H}{I}-to-dust ratio varies significantly in 3D, suggesting that a spatially constant dust-to-gas conversion may be a strong simplification. Extending this framework to include CO-traced $\rm H_{2}$ would provide a more complete picture of the hydrogen-to-dust ratio, and may enable such simulations to adopt a more realistic gas distribution.

In studies of kinematically resolved emitting gas, velocity separation is used a proxy for separation in physical space. Galactic rotation curves are often assumed in order to map the gas velocity to a physical distance. In theory, this highly resolved map of LOS gas velocity allows us to localise any velocity resolved emission in 3D space by tracing the LOS towards the emitting structure and identifying voxels along the LOS whose reconstructed velocity matches. Beyond this, MGVI and other IFT techniques can be used with this velocity map to convert any PPV (position/position/velocity) dataset within 1.25~kpc to a truly 3D PPP (position/position/position) dataset.

Our local/distant sky decomposition could also be used to inform further advancements in the \citet{soding25} gas map method, and our reconstructed gas-to-dust correlations could be used as a starting point for future joint dust/gas reconstructions.

Our conversions between the \citet{edenhofer23} map and dust mass densities have been carried out using a single dust model, at a single $R(V) = 3.1$. It has been shown that $R(V)$ can vary between around 2.0 and 4.0 in our volume of interest (\citet{zhang23_rv}, \citet{zhang25}, \citet{zhang25b}), which would change the conversion of extinction to dust density. For the purposes of this work we keep to our constant $R(V)=3.1$ assumption, but expect these advances in the 3D structure of $R(V)$ to lend themselves to even more precise conversions in the near future.

In this work we were fundamentally limited in reconstruction resolution by the size of MGVI problem that we could fit onto a single GPU. Both datasets used are available in higher resolution, with HI4PI going up to HEALPix $N_{\rm side} = 1024$ and the dust map up to $N_{\rm side} = 256$. Upcoming advancements in multi-GPU functionality in \texttt{NIFTy} will allow us to increase the resolution of this reconstruction, likely improving the quality of morphological matching between dust and gas structures towards the edge of the volume.

This method can also be developed further to include emission from CO, tracing the molecular hydrogen as well as the \ion{H}{I} (as was done in \citet{soding25} using data from \citet{dame01} and \citet{dame22}). This will provide stronger constraints on the velocity structure through further morphological matching between CO and dust structures, and in turn better constraints on gas densities. Another unexplored dataset which could be included is that of H$\alpha$ from the Wisconsin H-alpha Mapper (WHAM) \citet{haffner03}. This is a kinematic survey tracing the ionized gas. \citet{mccallum25} provides a 3D map of local H$\alpha$ emission, which could be used to pass further information on the velocity structure to the \ion{H}{I} and CO reconstruction. 

The extent of our map is also limited by the extent of the dust map. A larger dust-mapped volume would increase the quality of our local/distant sky decomposition and thus the fidelity of the local reconstruction. Upcoming Gaia data and further code advances in \texttt{NIFTy} (like multi-GPU functionality) will continue to improve both the quality and extent of the dust maps.

\section{Conclusions}

In this work we have developed and applied a new method for reconstructing the local atomic hydrogen in 3D, producing a velocity-resolved map of \ion{H}{I} emission within 1.25~kpc of the Sun. By combining the HI4PI 21-cm survey with the 3D dust map of \citet{edenhofer23} in the Information Field Theory framework, and performing inference with MGVI, we infer the local \ion{H}{I} density, radial velocity field, and effective line-width structure. A key aspect of the method is the inclusion of a remainder-sky component, which allows emission from beyond the mapped volume to be separated from genuinely local gas.

The resulting reconstruction provides a good match to the observed HI4PI data, with a reduced $\chi^{2}$ of 1.3 in the restricted velocity range $|V_{\rm LOS}|<75~\rm km~s^{-1}$. We find that approximately 50\% of this emission originates within 1.25~kpc, rising to 78\% at $|b|>30^{\circ}$. The inferred local sky follows much of the morphology seen in the dust map, while being systematically smoother and less sharply structured, as expected for the atomic phase of the ISM. The recovered vertical density profile is in reasonable agreement with the \citet{dickey90} description of local \ion{H}{I}, and the inferred \ion{H}{I}-to-dust ratios are seen to decrease above dust column densities of around $10^{-5} ~\rm g~cm^{-2}$, consistent with the expected transition from atomic to molecular hydrogen in the densest clouds \citep{krumholz09}.

Our reconstruction also yields a 3D LOS velocity field. This field shows the expected large-scale quadrupolar pattern from Galactic rotation, together with distinct regional deviations associated with local structures. Comparison to independent kinematic tracers shows that the inferred velocities are physically meaningful. The agreement is strongest for the sample of masers, which most directly traces the gas, while the match to the young stellar object clusters is weaker but still consistent once their additional dispersion relative to the gas is taken into account. These results suggest that the map could be useful for interpreting other velocity-resolved datasets in physical space.

The synthetic data tests provide support for the method. In a setup which includes both local emission and contributions from outside the 1.25~kpc volume, we recover the local density and velocity structure to a high degree of accuracy, and achieve a clean decomposition between local and distant emission, with only limited near/far confusion near the edge of the reconstructed volume.

The non-circular velocity dispersions and associated kinetic energy densities inferred here are broadly consistent with those reported by \citet{soler25}. This agreement supports the conclusion that the turbulent motions measured on these scales are not achievable through only supernova feedback, and may point to other sources of velocity dispersion, as discussed by \citet{abboudeh26}.

This paper represents a methodological advance over earlier kinetic-tomography work \citep{kt1} and the HOG-based reconstruction of \citet{soler25}. Its main strengths are the improved spatial resolution enabled by modern 3D dust maps and HI4PI, as well as the ability allowed by MGVI to infer tens of millions of parameters. At the same time, some difficulties remain such as line-of-sight confusion and dependence on smoothing assumptions. Future higher velocity resolution \ion{H}{I} surveys will be essential for further improving such reconstructions.

Several limitations remain. The reconstruction uses a fixed spin temperature for the \ion{H}{I} radiative transfer, a fixed dust model and $R(V)=3.1$ for converting extinction to dust mass density, and the limited extent of the underlying dust map. In addition, the limited resolution imposed by current single-GPU MGVI calculations makes the outer regions of the volume more susceptible to confusion between truly local gas and emission arising just beyond 1.25~kpc. Structures such as the Magellanic Stream are also difficult to absorb cleanly into the distant remainder model, particularly where they overlap the velocity range of local gas.

Despite these limitations, this work demonstrates that IFT techniques can combine 3D dust maps with spectrally resolved \ion{H}{I} data to recover a physically meaningful, velocity-resolved picture of the local ISM. The resulting map is not just a demonstration of the method, but a usable 3D data product for studies of nearby Galactic structure, for placing PPV emission in physical space, and for comparison to other tracers of the local ISM. With improved 3D dust maps, larger mapped volumes, inclusion of additional tracers such as CO and H$\alpha$, and upcoming multi-GPU capabilities in \texttt{NIFTy}, this method should enable increasingly detailed reconstructions of the nearby Milky Way.

\section*{Acknowledgements}

Funded by the European Union. Views and opinions expressed are however those of the author(s) only and do not necessarily reflect those of the European Union or the European Research Council Executive Agency. Neither the European Union nor the granting authority can be held responsible for them. This work is supported by ERC grant (mw-atlas, 101166905).

JDS is funded by the Austrian Science Fund (Fonds zur Förderung der wissenschaftlichen Forschung, FWF) under Grant DOI 10.55776/PAT6169824.

We greatly thank Theo O'Neill and Catherine Zucker for sharing results and data from their in prep. work on local IVCs, helping to confirm our recovery of many major clouds at high latitude.

We also thank Fabian Polnitzky for the fruitful discussions and comments.

\section*{Data Availability}

3D grids of reconstructed \ion{H}{I} density, LOS velocity, line width and the remainder datacube are available for download at \url{https://doi.org/10.5281/zenodo.21262835}. We include posterior mean maps on both spherical and Cartesian grids, and our 10 posterior samples on the spherical grid.



\bibliographystyle{mnras}
\bibliography{biblio} 




\appendix

\section{Radially Integrated Shells}

\begin{figure*}
    \centering
    \includegraphics[width=0.9\textwidth]{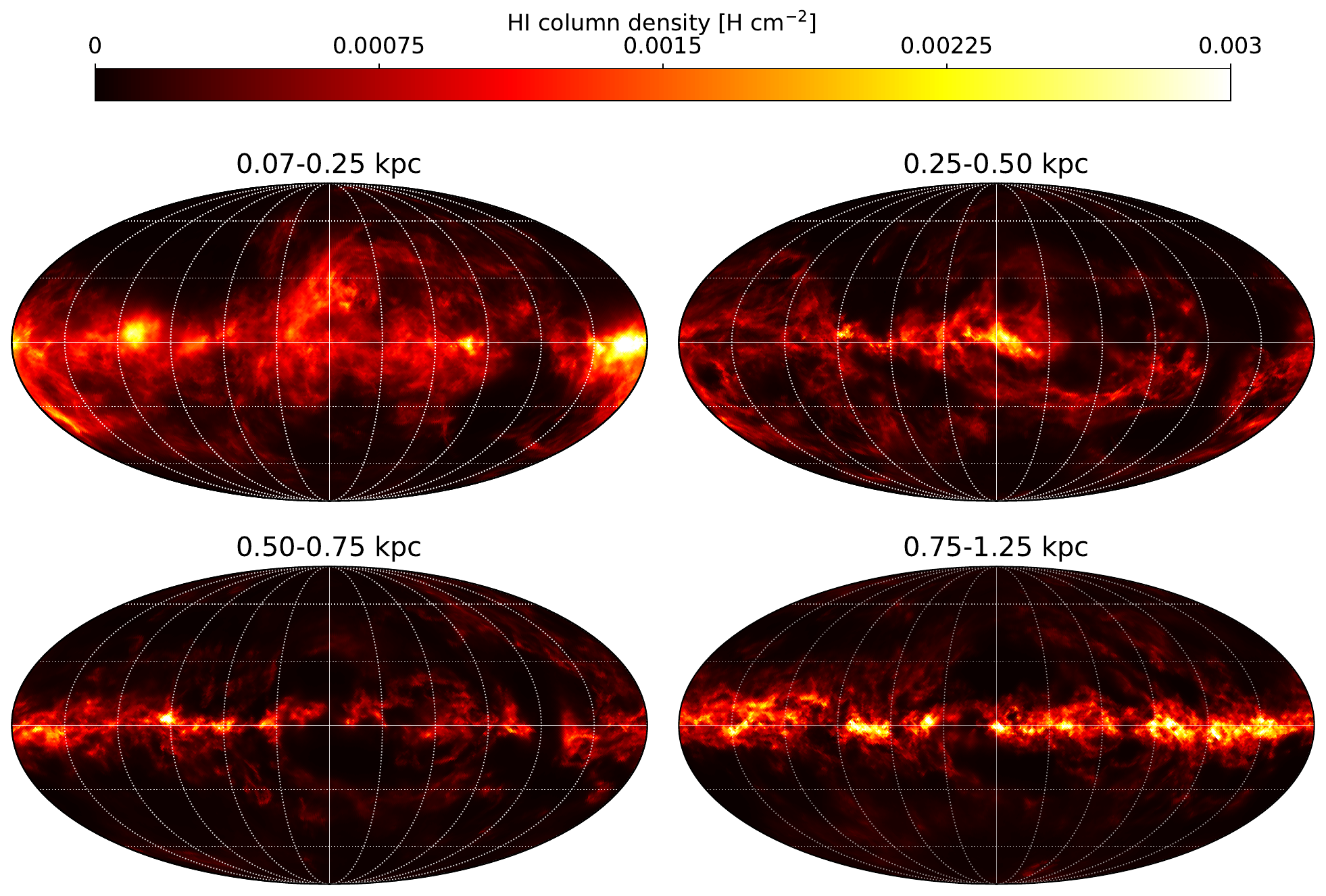}
    \caption{A veiw towards the Galactic centre ($\ell = 0^{\circ}$) in radially integrated shells of \ion{H}{I} density. These do not included radiative transfer effects, and are simply the total \ion{H}{I} column density through the following radius ranges: 70-250~pc, 250-500~pc, 500-750~pc, and 750-1250~pc.}
    \label{fig:radialshells}
\end{figure*}

\begin{figure*}
    \centering
    \includegraphics[width=0.9\textwidth]{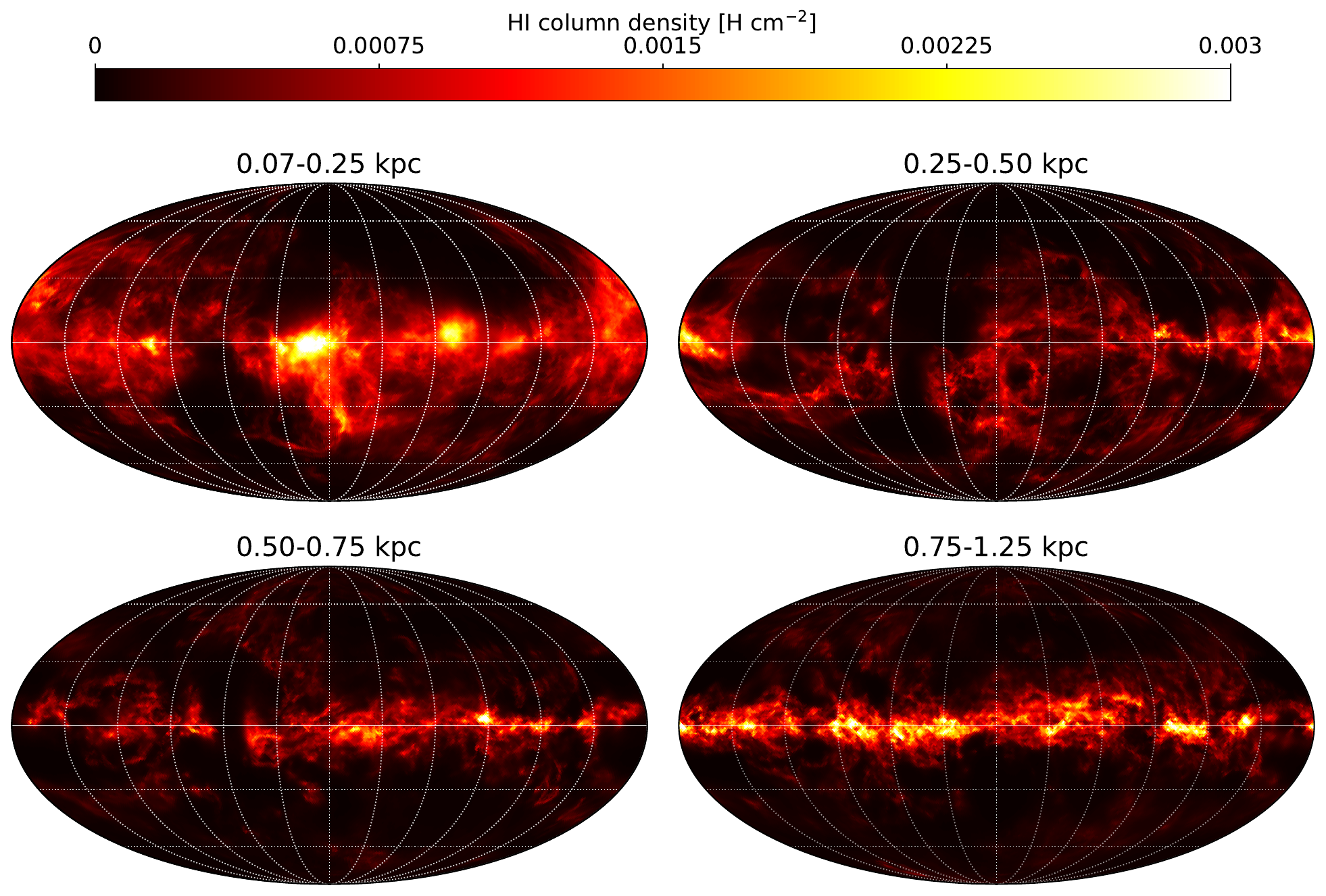}
    \caption{The same as figure~\ref{fig:radialshells}, but centred on the Galactic anti-centre at $\ell = 180^{\circ}$.}
    \label{fig:radialshellsanti}
\end{figure*}

To allow the reader to pick out clouds in particular volumes of interest, we also include some radially integrated shells of the \ion{H}{I} density grid. Figure~\ref{fig:radialshells} shows our \ion{H}{I} map centred on the Galactic centre between 70-250pc, 250-500pc, 500-750pc and 750-1250pc. The same shells are shown in figure~\ref{fig:radialshellsanti}, but with the maps centred on the Galactic anti-centre. 

\section{Map Cross Sections and $z$-restricted intervals}

\begin{figure*}
    \centering
    \includegraphics[width=1.0\textwidth]{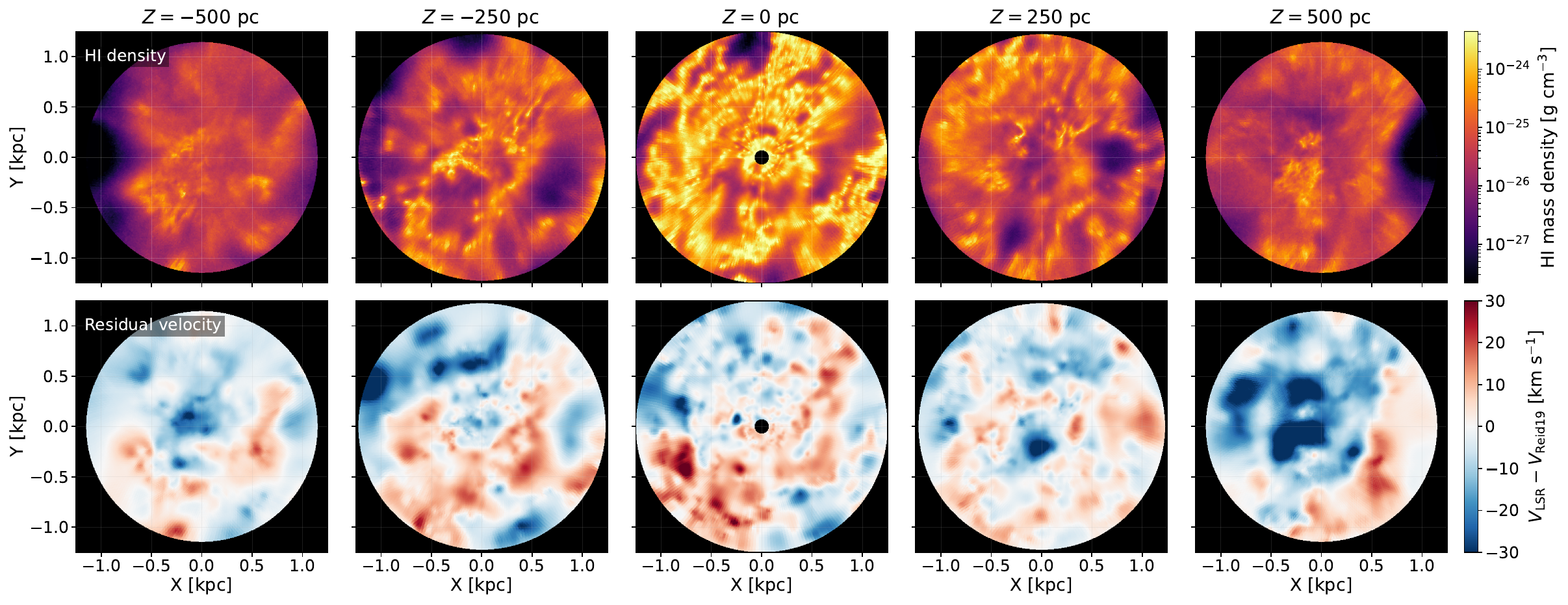}
    \caption{Vertical slices through the reconstructed local H I density and velocity field. The top row shows H I mass density in five planes parallel to the Galactic midplane at \(Z=-500,-250,0,+250,+500\) pc. The bottom row shows the corresponding reconstructed residual line-of-sight velocity, \(V_{\rm LSR}-V_{\rm Reid19}\), on the same spatial grid. It is worth noting that slices of the velocity grid at higher values of $Z$ probe more of the vertical gas velocities, while the $Z=0~\rm pc$ slice is probing only motion in the axis of the plane of the Galaxy.}
    \label{fig:crosssections} 
\end{figure*}

To complement the integrated face-on projections shown in the main text, Fig.~\ref{fig:crosssections} shows cross-sections through the reconstructed posterior-mean \ion{H}{I} density field and posterior-mean velocity field (with Galactic rotation subtracted). As mentioned in the main text, the empty circular region in the $Z=0~\rm pc$ slice reflects the inner volume not covered by the \citet{edenhofer23} dust map. These cross sections were evaluated on a Cartesian grid spanning the $2.5~\rm kpc^{3}$ and $512^{3}$ grid cells. These cross sections thus have an effective depth of 4.9~pc.

\begin{figure*}
    \centering
    \includegraphics[width=1.0\textwidth]{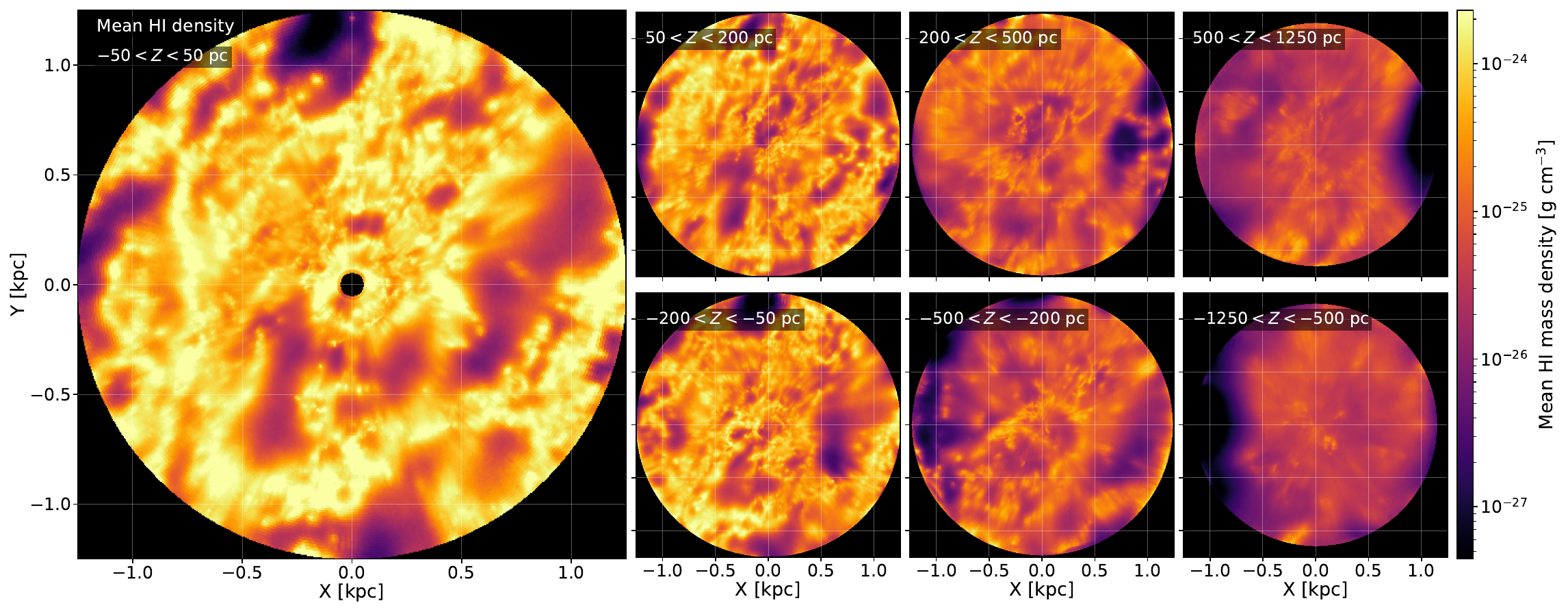}
    \includegraphics[width=1.0\textwidth]{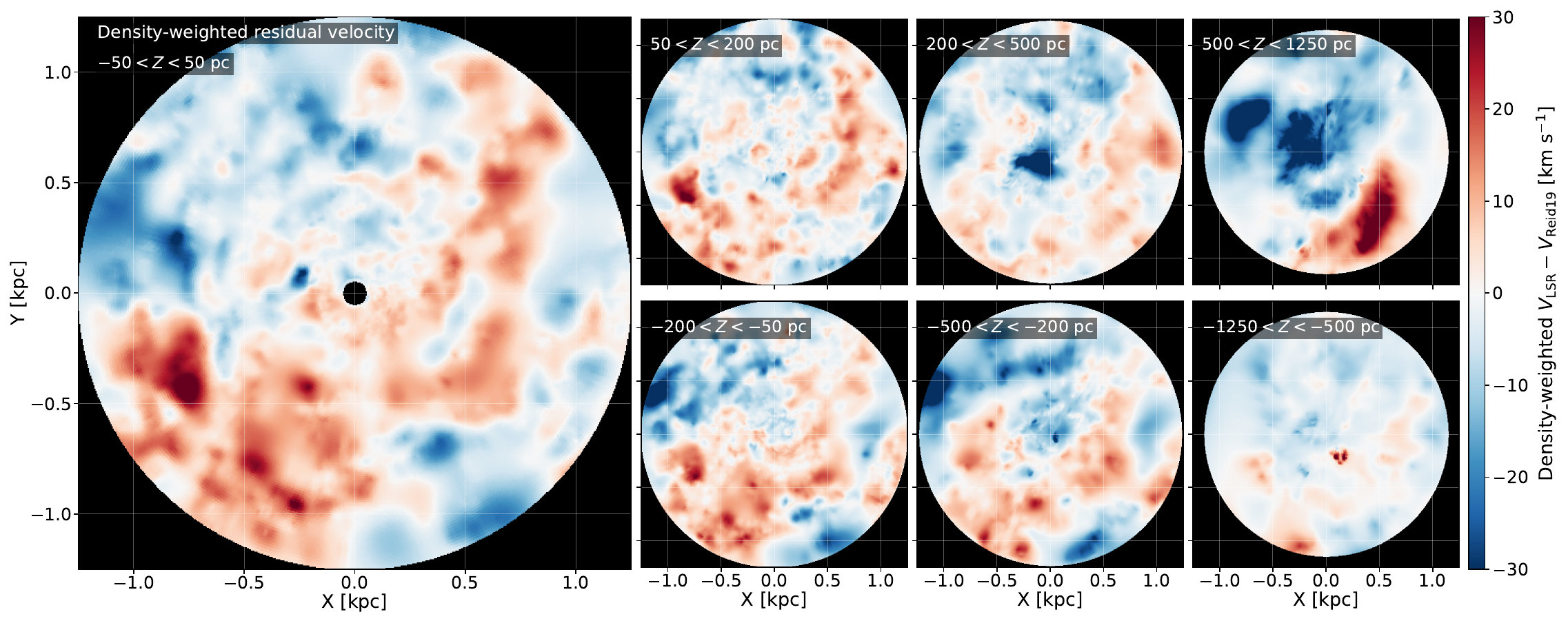}
    \caption{\ion{H}{I} structure in finite $Z$-intervals. Top: \ion{H}{I} column density integrated through each labelled $Z$ interval. Bottom: H I-density-weighted line-of-sight velocity over the same intervals. This is residual velocity after Galactic rotation from \citet{reid19} has been subtracted. The large panel shows the mid-plane slab, $-50<Z<50$ pc, while the smaller panels separate positive and negative heights above and below the Galactic plane.}
    \label{fig:zintervals} 
\end{figure*}

In order to help the reader identify individual structures at specific $z$-heights, in figure~\ref{fig:zintervals} we include views of the \ion{H}{I} density spanning various restricted intervals in the $z$-axis. For each $z$-interval we include both a total \ion{H}{I} column density, and a \ion{H}{I} density weighted velocity in the frame of $\rm V_{LSR} - V_{Reid19}$. The large panels on the left show the slab which straddles the midplane, with various heights above and below the midplane shown in the smaller panels to the right.

\section{Table of Priors}

Table~\ref{priorstable} shows our prior choices for our main reconstruction.

\begin{table*}
\centering
\caption{Table of used priors.}
\begin{tabular}{lccccc}
\hline\hline
Name & Distribution & Mean & Standard deviation & Units & Degrees of Freedom \\
\hline
\multicolumn{6}{c}{Local $\log({\rm \ion{H}{I}/dust})$ field} \\
\hline
Latent multigrid coefficients & Normal & 0 & 1.0 & -- & 11,047,476 \\
Mean $\log({\rm \ion{H}{I}/dust})$ offset & Normal & 4.7 & 0.2 & -- & 1 \\
Mat\'ern variance & Log-normal & 0.5 & 0.1 & -- & 1 \\
Mat\'ern lengthscale & Log-normal & 0.3 & 0.1 & kpc & 1 \\
Mat\'ern slope & Normal & 3.5 & 0.5 & -- & 1 \\
\hline
\multicolumn{6}{c}{Local LOS velocity field} \\
\hline
Latent multigrid coefficients & Normal & 0 & 1.0 & -- & 11,047,476 \\
Mean velocity offset & Normal & 0.01 & 1.0 & km s$^{-1}$ & 1 \\
Mat\'ern variance & Log-normal & 5.0 & 2.0 & (km s$^{-1}$)$^{2}$ & 1 \\
Mat\'ern lengthscale & Log-normal & 0.15 & 0.05 & kpc & 1 \\
Mat\'ern slope & Normal & 3.5 & 0.5 & -- & 1 \\
\hline
\multicolumn{6}{c}{Local line-width field} \\
\hline
Latent multigrid coefficients & Normal & 0 & 1.0 & -- & 11,047,476 \\
Mean log line-width offset & Normal & 1.1 & 0.05 & $\log(\mathrm{km\,s^{-1}})$ & 1 \\
Mat\'ern variance & Log-normal & 0.05 & 0.01 & $\log(\mathrm{km\,s^{-1}})^2$
 & 1 \\
Mat\'ern lengthscale & Log-normal & 0.3 & 0.1 & kpc & 1 \\
Mat\'ern slope & Normal & 3.5 & 0.5 & -- & 1 \\
\hline
\multicolumn{6}{c}{Distant remainder field} \\
\hline
Latent sky--velocity coefficients & Normal & 0 & 1.0 & -- & 3,276,600 \\
Longitude latent field for $H_{\rm thick}(\ell)$ & Normal & 0 & 1.0 & -- & 32 \\
Longitude latent field for $H_{\rm thin}(\ell)$ & Normal & 0 & 1.0 & -- & 32 \\
Longitude latent field for $S_{\rm thin}(\ell)$ & Normal & 0 & 1.0 & -- & 32 \\
Longitude latent field for $B_0(\ell)$ & Normal & 0 & 1.0 & -- & 32 \\
Global log-intensity offset & Normal & 0.2 & 0.02 & -- & 1 \\
Sky variance & Log-normal & 0.11 & 0.05 & -- & 1 \\
Sky lengthscale & Log-normal & 1.9 & 0.4 & rad & 1 \\
Sky power-spectrum slope & Normal & 3.25 & 0.4 & -- & 1 \\
Velocity smoothing $\sigma_v$ & Log-normal & 4.6 & 1.0 & km s$^{-1}$ & 1 \\
Thick-disc scale height $H_{{\rm thick},0}$ & Log-normal & 16.0 & 4.0 & deg & 1 \\
Thick-disc longitude smoothing & Log-normal & 0.025 & 0.01 & rad & 1 \\
Thick-disc longitude modulation amplitude & Log-normal & 0.01 & 0.005 & -- & 1 \\
Thin-disc scale height $H_{{\rm thin},0}$ & Log-normal & 2.0 & 0.7 & deg & 1 \\
Thin-disc longitude smoothing & Log-normal & 0.011 & 0.005 & rad & 1 \\
Thin-disc scale-height modulation amplitude & Log-normal & 0.032 & 0.015 & -- & 1 \\
Thin-disc amplitude longitude smoothing & Log-normal & 0.003 & 0.002 & rad & 1 \\
Thin-disc amplitude modulation & Log-normal & 0.42 & 0.15 & -- & 1 \\
Global thin-disc weight & Log-normal & 1.2 & 0.3 & -- & 1 \\
Mid-plane shift longitude smoothing & Log-normal & 0.012 & 0.006 & rad & 1 \\
Mid-plane shift amplitude & Log-normal & 0.09 & 0.05 & deg & 1 \\
\hline
\label{priorstable}
\end{tabular}
\end{table*}


\bsp	
\label{lastpage}
\end{document}